\documentclass[10pt]{iopart} 
\usepackage[english]{babel}
\usepackage{iopams} 

\newif\ifnight
\nightfalse
\expandafter\let\csname equation*\endcsname\relax
\expandafter\let\csname endequation*\endcsname\relax
\usepackage{amsmath}
\usepackage{amssymb}
\usepackage{color}
\usepackage{geometry}
\usepackage{ifthen} 
\usepackage[pdftex]{graphicx}
\usepackage[pdftex]{hyperref} 
\ifnight
\usepackage{xcolor}
\pagecolor[rgb]{0.2,0.2,0.2} %black
\color[rgb]{0.7,0.7,0.7} %grey
\hypersetup{colorlinks=true,
	linkcolor=yellow,
	citecolor=yellow,
	urlcolor=yellow,
	pdfauthor={Paul Heinrich IPP, paul.heinrich@ipp.mpg.de},
	pdfsubject={Heinrich 2024 Radiated energy fraction of SPI-induced disruptions at ASDEX Upgrade},
	pdftitle={Radiated energy fraction of SPI-induced disruptions at ASDEX Upgrade}
}
\else
\hypersetup{colorlinks=true,
	linkcolor=blue,
	citecolor=blue,
	urlcolor=blue,
	pdfauthor={Paul Heinrich IPP, paul.heinrich@ipp.mpg.de},
	pdfsubject={Heinrich 2024 Radiated energy fraction of SPI-induced disruptions at ASDEX Upgrade},
	pdftitle={Radiated energy fraction of SPI-induced disruptions at ASDEX Upgrade}
}
\fi
\DeclareGraphicsExtensions{.pdf,.png}

\usepackage[numbers,sort&compress]{natbib}
\usepackage{hypernat}
\def\newblock{\hskip .11em plus .33em minus .07em}

\usepackage{bmpsize}
\usepackage{caption}
\usepackage{mwe}
\usepackage{subfigure}
\usepackage{lipsum}
\usepackage{longtable}
\usepackage{hhline}
\usepackage[normalem]{ulem}
\usepackage{tikz}
\usepackage{makecell}

\usepackage{todonotes}
\newcommand*\circled[1]{\tikz[baseline=(char.base)]{%
            \node[shape=circle,draw,inner sep=1.5pt] (char) {#1};}}

\definecolor{pythonblue}{RGB}{31,119,180}
\definecolor{pythonorange}{RGB}{255,127,14}
\definecolor{pythongreen}{RGB}{44,160,44}
\definecolor{pythoncoolwarmblue}{RGB}{117,151,246}
\definecolor{pythoncoolwarmbrown}{RGB}{237,209,194}
\definecolor{pythoncoolwarmorange}{RGB}{242,146,116}

\usepackage[export]{adjustbox}

\usepackage{xspace}
\newcommand{\Ip}{\ensuremath{I_\text{P}}\xspace}
\newcommand{\Wpl}{\ensuremath{\text{W}_\text{plasma}}\xspace}
\newcommand{\Wth}{\ensuremath{\text{W}_\text{th}}\xspace}
\newcommand{\Wmag}{\ensuremath{\text{W}_\text{mag}}\xspace}
\newcommand{\Wcoupl}{\ensuremath{\text{W}_\text{c}}\xspace}
\newcommand{\Wexheat}{\ensuremath{\text{W}_\text{heat}}\xspace}
\newcommand{\Wrad}{\ensuremath{\text{W}_\text{rad}}\xspace}

\newcommand{\frad}{\ensuremath{\text{f}_\text{rad}}\xspace}
\newcommand{\fc}{\ensuremath{\text{f}_\text{c}}\xspace}
\newcommand{\fth}{\ensuremath{\text{f}_\text{th}}\xspace}

\newcommand{\tFL}{\ensuremath{\text{t}_\text{FL}}\xspace}

\newcommand{\tCQend}{\ensuremath{\text{t}_\text{CQ-end}}\xspace}
\newcommand{\trecovery}{\ensuremath{\text{t}_\text{recovery}}\xspace}
\newcommand{\tend}{\ensuremath{\text{t}_\text{end}}\xspace}

\newcommand{\dt}{\text{d}t}
\newcommand{\vperp}{\ensuremath{\text{v}_\perp}\xspace}

\newcommand{\VPEState}{\texttt{VPEState}}
\newcommand{\DCSVPEState}{\texttt{DCS/\VPEState}}

\begin{document}
\hyphenation{ASDEX}

\title{Radiated energy fraction of SPI-induced disruptions at ASDEX Upgrade}
\author{P.~Heinrich$^{1, 2}$, G.~Papp$^1$, S.~Jachmich$^3$, J.~Artola$^3$, M.~Bernert$^1$, P.~de~Marn\'e$^1$, M.~Dibon$^3$, R.~Dux$^1$, T.~Eberl$^1$, J.~Hobirk$^{1}$, M.~Lehnen$^3$\footnote{Deceased}, T.~Peherstorfer$^4$, N.~Schwarz$^3$, U.~Sheikh$^5$, B.~Sieglin$^1$, J.~Svoboda$^6$, the ASDEX Upgrade Team$^a$ and the EUROfusion Tokamak Exploitation Team$^{b}$}
\address{$^1$Max Planck Institute for Plasma Physics, Boltzmannstr. 2, D-85748 Garching, Germany}
\address{$^2$Technical University of Munich, TUM School of Natural Sciences, Physics Department, James-Franck-Str. 1, D-85748 Garching, Germany}
\address{$^3$ITER Organization, Route de Vinon-sur-Verdon, CS 90 046 13067 St.~Paul-lez-Durance, France}
\address{$^4$Institute for Applied Physics, Wiedner Hauptstr. 8-10/134, 1040 Wien, Austria}
\address{$^5$Ecole Polytechnique F\'{e}d\'{e}rale de Lausanne (EPFL), Swiss Plasma Center (SPC), CH-1015 Lausanne, Switzerland}
\address{$^6$Institute of Plasma Physics of the CAS, CZ-18200 Praha 8, Czech Republic}
\address{$^a$See the author list of \href{https://doi.org/10.1088/1741-4326/ad249d}{H.~Zohm~\etal 2024 Nucl. Fusion}} 
\address{$^b$See the author list of \href{https://doi.org/10.1088/1741-4326/ad2be4}{E.~Joffrin~\etal 2024 Nucl. Fusion}}

\date{\today}

\begin{abstract}
    Future large tokamaks will operate at high plasma currents and high stored plasma energies. To ensure machine protection in case of a sudden loss of plasma confinement (major disruption), a large fraction of the magnetic and thermal energy must be radiated to reduce thermal loads.
    The disruption mitigation system for ITER is based on massive material injection in the form of shattered pellet injection (SPI). To support ITER, a versatile SPI system was installed at the tokamak ASDEX Upgrade~(AUG). The AUG SPI features three independent pellet generation cells and guide tubes, and each was equipped with different shatter heads for the 2022 experimental campaign. We dedicated over 200 plasma discharges to the study of SPI plasma termination, and in this manuscript report on the results of bolometry (total radiation) analysis.
    We found, that the amount of neon inside the pellets is the dominant factor determining the radiated energy fraction~(\frad). Large and fast fragments, produced by the $12.5^{\circ}$ rectangular shatter head, lead to somewhat higher values of \frad compared to the $25^{\circ}$ circular or rectangular heads.
    This effect is strongest for neon content of \mbox{$\lesssim 3 \times 10^{20}$} neon atoms ($\textrm{f}_\textrm{neon} \lesssim 1.25\%$ neon) injected, where a lower normal velocity component (larger fragments) seems slightly beneficial. While full-sized, 8~mm diameter, 100\% deuterium ($\textrm{D}_2$) pellets lead to a disruption, the 4~mm or shortened 8~mm pellets of 100\% $\textrm{D}_2$ did not. The disruption threshold for 100\% $\textrm{D}_2$ is found to be around \mbox{$1\times 10^{22}$} deuterium molecules inside the pellet. While the radiated energy fraction of non-disruptive SPI is below 20\%, this is increased to 40\% during the TQ and VDE phase of the disruptive injections. For deuterium--neon--mix pellets, \frad-values of \mbox{$\leq 90$\%} are observed, and the curve saturates around 80\% already for 10\% neon mixed into the 8~mm pellets (\mbox{$2\times 10^{21}$} neon atoms).
\end{abstract}

\submitto{\NF}

\maketitle
\ioptwocol 

\section{Introduction \label{sec:Introduction}}
Plasma terminating disruptions pose a major challenge for tokamaks with high plasma current and stored energy~\cite{Hender_2007_ITER_Physics, Hollmann_2011_PFCdisruptions, Matthews_2016}. 
An efficient disruption mitigation system~(DMS) should address three main challenges:
minimise the thermal loads on the plasma facing components~(PFCs), minimise eddy and halo currents which cause large mechanical loads, while suppressing the formation of a large runaway electron~(RE) beam~\cite{Hender_2007_ITER_Physics}. In present day tokamaks, disruption mitigation is typically realised using massive material injection~(MMI), either via massive gas injection~(MGI) or shattered pellet injection~(SPI). High-Z material inside the injection mix helps spread the energy over a larger surface area via radiation, reducing the localised thermal loads~\cite{Lehnen_2015_ITER, Moscheni_2020_radDEMO}. The mechanical loads can be reduced by tailoring the duration of the current quench~(CQ) through the injected material composition~\cite{Lehnen_2015_ITER, Strauss_2018_fastCQ} or early injections in cases of vertical displacement events~(VDEs)~\cite{Schwarz_2023, Schwarz_2024_PhD}. The formation of a RE beam is hoped to be suppressed by increasing the free electron density in the plasma core~\cite{Breizman_2019}, hence the material assimilation (of low Z material -- namely protium/deuterium) is a key factor~\cite{Lehnen_2023_FEC, Martin-Solis_2017_RE, Pautasso_2020}.

The SPI technique was first tested at \mbox{DIII-D} in 2009~\cite{Commaux_2010, Baylor_2019_SPI-DIII-D-JET}, and over time more tokamaks, such as JET~\cite{Baylor_2019_SPI-DIII-D-JET}, KSTAR~\cite{Park_2020_KSTAR-SPI, Park_2021_KSTAR-SPI-experiments}, HL-2A~\cite{Xu_2019_HL-2A-SPI}, J-TEXT~\cite{Li_2018_J-TEXT-SPI-design,Li_2021_J-TEXT-SPI-MGI}, and EAST~\cite{Zhao_2023_EAST-SPI-control-system, Yuan_2023_EAST-SPI} followed.

In late 2021, a highly flexible SPI system was installed at ASDEX Upgrade~(AUG)~\cite{Dibon_2023_SPI, Jachmich_2023_EPS,Heinrich_2024_SPI_Lab, Heinrich_2025_PhD} to provide further input for the design and optimisation of the ITER DMS~\cite{Lehnen_ITER_workshop_2021,Lehnen_2023_FEC} based on the SPI principle~\cite{Baylor_2010_DMS_SPI}. In this paper we discuss the evolution of total radiated energy~(\Wrad) and the radiated energy fraction~(\frad) as a function of pellet and injection parameters in different plasma scenarios, based on the bolometry analysis of the 2022 AUG SPI experimental campaign.

In section~\ref{sec:Experimental_setup} the experimental setup for the 2022 experimental campaign of AUG is introduced. The derivation of the radiated power measurements is discussed in section~\ref{sec:Radiated_power}. In section~\ref{sec:frad} the calculation of the radiated energy fraction and other derived quantities is presented. The experimental values of \Wrad and \frad are presented in section~\ref{sec:Wrad_frad_values} with the effect of the different shatter head geometries in section~\ref{sec:Effect_geometry}.

\section{Experimental setup \label{sec:Experimental_setup}}
\subsection{Shattered pellet injector on AUG\label{sec:injector_on_AUG}}
The AUG SPI is a triple-barrel system, with three independent guide tubes and three different shatter heads~\cite{Dibon_2023_SPI, Heinrich_2024_SPI_Lab}. Three pellets, made of mixtures of deuterium and neon (as well as 100\% deuterium or neon) can be generated and launched simultaneously, potentially allowing the study of multi-injection scenarios. A detailed description of the SPI system is given in the paper by \mbox{Dibon~\etal~\cite{Dibon_2023_SPI}} and the laboratory commissioning\footnote{Animation video of the laboratory setup available at https://datashare.mpcdf.mpg.de/s/DlMzGcWnZwoHMjq~or~\cite{Heinrich_2024_SPI_Lab}.} by \mbox{Heinrich~\etal~\cite{Heinrich_2024_SPI_Lab, Heinrich_2025_PhD}}.

\begin{figure*}[htb]
	\centering
	\includegraphics[width=\textwidth]{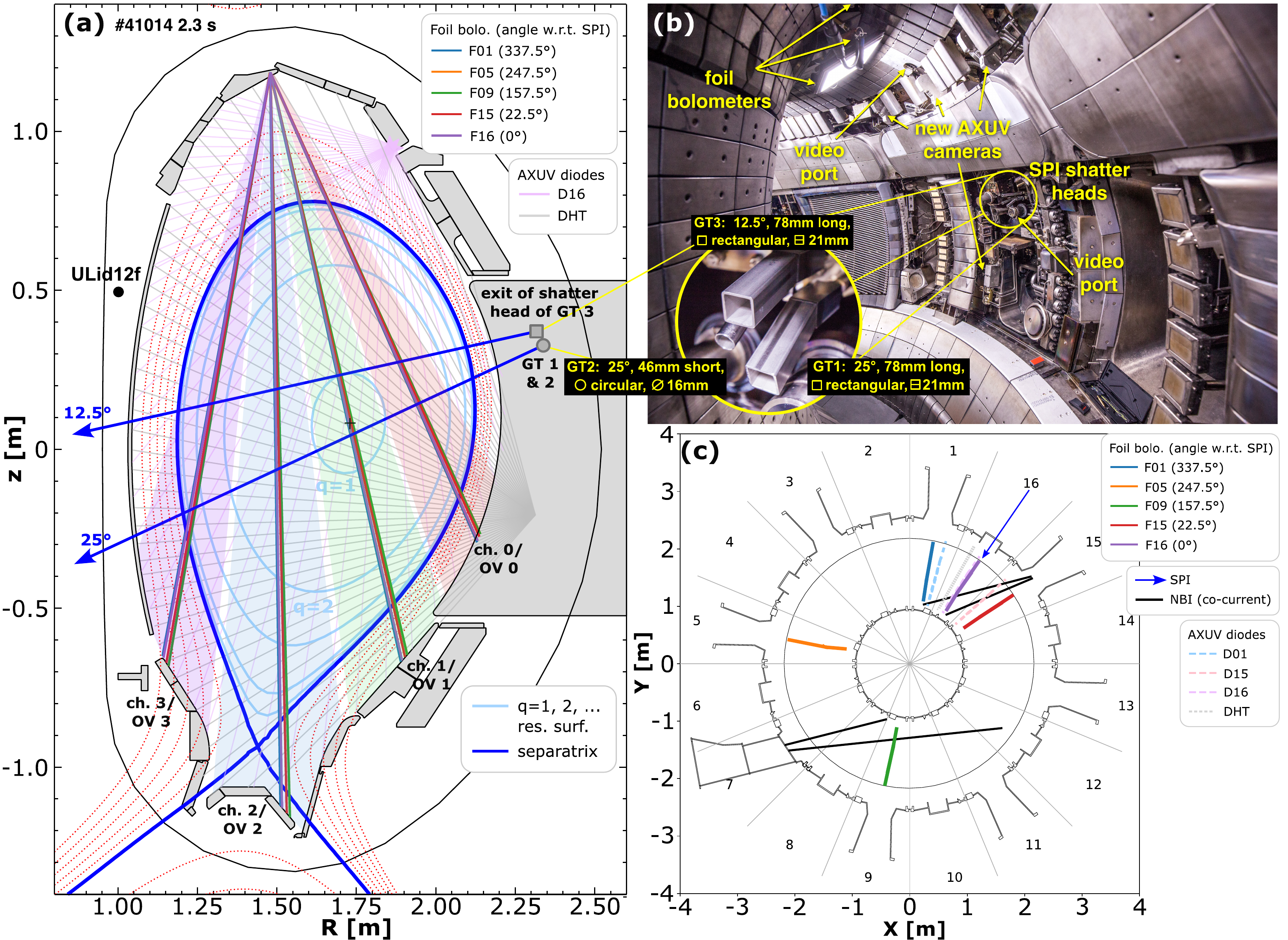}
	\caption{Geometry of SPI and bolometry measurements in AUG. (a) Poloidal cross section. The observation volumes of the foil bolometers are indicated with the shaded volumes with channel numbers 0 (right) to 3 (left). (b) In-vessel picture with the insert showing the shatter head configuration for the 2022 campaign. (c) Toroidal cross-section (top-down view). \label{fig:exp_setup}}
\end{figure*}

\begin{table*}[htb]
        \caption{Geomety of shatter heads installed for the 2022 AUG campaign. The miter bend angle of the shatter head is denoted with $\alpha$. Note that for the circular shatter head the effective shatter angle may vary (see figure~\ref{fig:head_comparison}).}
        \label{tab:shatter_heads}
        \centering
        \begin{tikzpicture}
            \node (table) [inner sep=0pt] {
            \begin{tabular}{l|c|l|l|l}
                Guide tube & $\alpha$ [$^{\circ}$] & head size [mm] & shape & description\\
                \hline
                \hline
                GT1 & 25 & $21\times 78$ (W $\times$ L) & rectangular, & collimated fragment plume;\\
                & & & long & smaller \& slower fragments\\
                \hline
                GT2  & 25 & $16\times 46$ (D $\times$ L) &  circular, & wide fragment plume;\\
                & & & short & smaller \& slower fragments\\
                \hline
                GT3 & 12.5 & $21 \times 78$ (W $\times$ L) & rectangular, & collimated fragment plume;\\
                & & & long & larger \& faster fragments\\
            \end{tabular}
            };
            \draw [rounded corners=.5em] (table.north west) rectangle (table.south east);
	\end{tikzpicture}
\end{table*}

\begin{figure*}[htb]
	\centering
	\includegraphics[width=\textwidth]{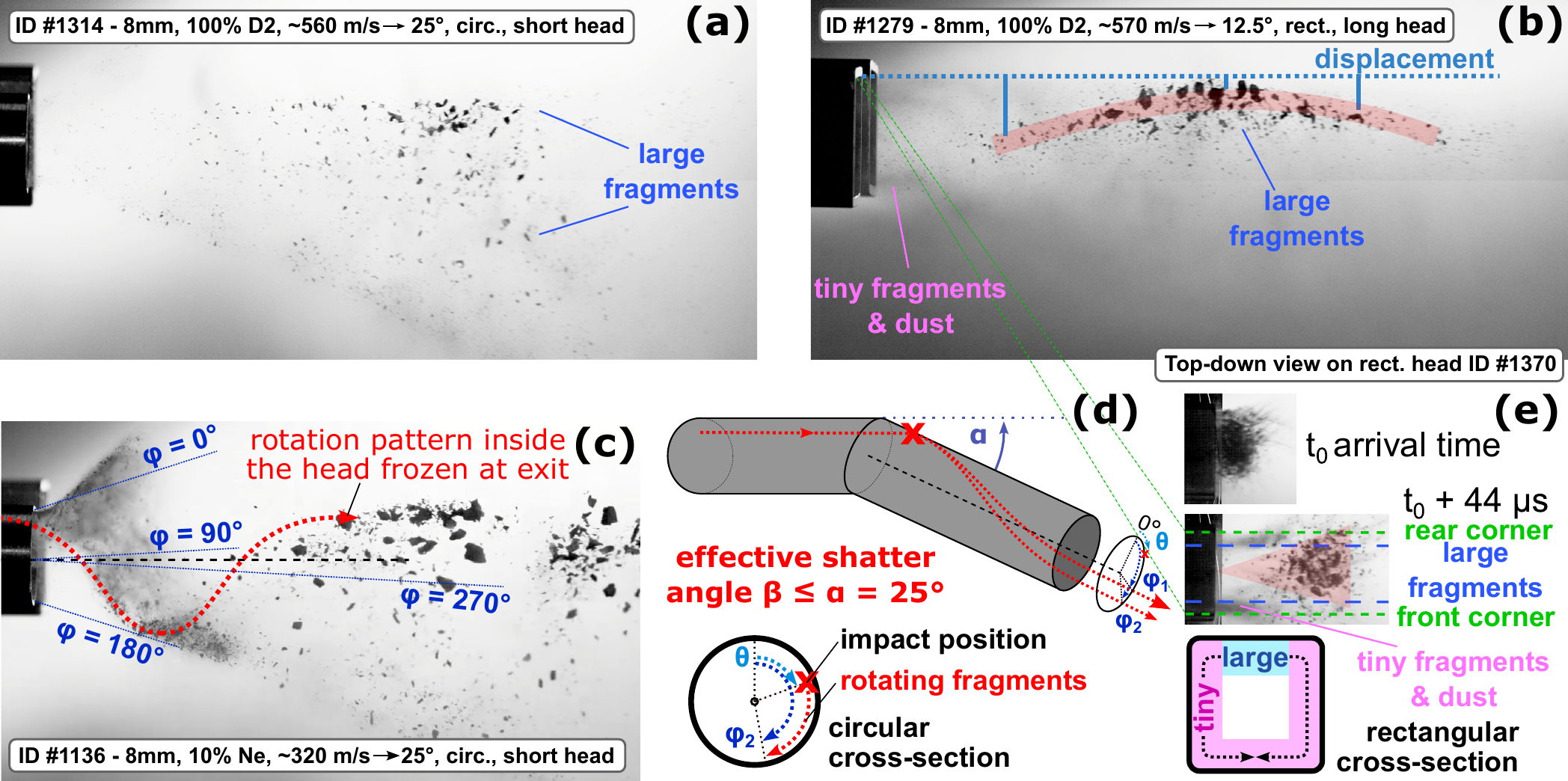}
	\caption{Comparison of the spatial distribution of the fragment plumes generated by different shatter heads. In~(a) a larger spread of the fragments is observed for the circular, $25^{\circ}$ shatter head compared to the more collimated fragment plume of the rectangular, $12.5^{\circ}$ head for similar pre-shattering pellet velocity in~(b). In~(c)~and~(d) the geometrical effects of the circular shatter head are shown. The effective shatter angle $\beta$ depends on the impact position ($\theta$) inside a circular shatter head~\cite{Peherstorfer_2022_fragmentation}, while for the rectangular head the shatter angle is equal to the miter bend angle $\alpha$. Additionally, the shards may receive perpendicular torque inside the circular head, which results in spiral patterns as shown in~(c,~d), also increasing the spread of the fragments. After leaving the exit of the shatter head, the spatial distribution is set by the position $\varphi$ at the time of exit, depending on the rotation inside the head. This is indicated by the red, dashed line in~(c,~d). In contrast, the ``top-down'' view for the rectangular shatter head in~(e) indicates, that the large main fragments leave the shatter head centrally on the ``shatter plane'' ($\theta = \varphi \approx 0^{\circ}$; top side) and only the tiny fragments and dust are pushed towards and around the corners ($\varphi > 0^{\circ}$). Hereby the fragments seem to have a minor triangular fragment plume, similar to the observations by T.~Gebhart~\etal~\cite{Gebhart_2021}. In the side view in~(b), this will result in the collimated, arched fragment plume: large fragments will have a minor displacement to the shatter plane, whereas small fragments will be pushed around the corners and appear further downwards (with larger displacement).\label{fig:head_comparison}}
\end{figure*}

Following extensive laboratory commissioning and the analysis of fast camera recordings of the resulting pellet fragment clouds~\cite{Peherstorfer_2022_fragmentation,Illerhaus_2024_SPI}, three different shatter heads (all with miter bends) were selected for the 2022 experimental campaign~\cite{Dibon_2023_SPI} (see figure~\ref{fig:exp_setup}(a, b) and table~\ref{tab:shatter_heads}). 
A short, circular cross-section, $25^\circ$ head is used for increased spatial spread of the fragments. Two long, rectangular cross-section heads were installed for better collimation. These are a ``matching pair'' with shatter angles of $12.5^\circ$ and $25^\circ$ respectively, which allows the more convenient matching of the normal impact velocity~(\vperp) -- the main factor for the pellet fragment size distribution. As $\sin(25^\circ) \approx 2\cdot \sin(12.5^\circ)$, different parallel velocity components -- hence fragment velocities~\cite{Peherstorfer_2022_fragmentation} -- are achieved while matching \vperp. Hereby, a higher normal impact velocity will typically result in smaller fragments~\cite{Parks_2016_model, Gebhart_2020_IEEE, Peherstorfer_2022_fragmentation, Gebhart_2021}. This allows to test the effects of the fragment size distributions and injection velocities on the disruptions independently.

Figure~\ref{fig:head_comparison} shows typical spatial distributions of the fragments (``side view'') for the $25^\circ$, circular, and shortened~(a,~c) and $12.5^\circ$, rectangular, and long~(b,~e) shatter head as observed during the commissioning phase. Overall the rectangular shatter head has a more collimated fragment plume -- with fragments leaving the top side (``shatter plane'') centrally and only the small fragments pushed towards and around the corners (compare figure~\ref{fig:head_comparison}(b)) as shown by the ``top-down'' video recordings (see figure~\ref{fig:head_comparison}(e)).
This will create an arched fragment plume as the large fragments leaving the head at the shatter plane show a low displacement w.r.t. the upper edge of the shatter head compared to the smaller fragments, that are pushed around the corners, hence showing a larger displacement. For the top-down view, the fragment plume of the rectangular head seems more spread out, with a slight triangular pattern indicated in figure~\ref{fig:head_comparison}(e) as also observed by T.~Gebhart~\etal~\cite{Gebhart_2021}. Compared to the so-called ``horn'' geometry (rectangular head with fixed height but widening width~\cite{Heinrich_2025_PhD}), the only difference is observed for the tiny fragments and dust, pushed all the way to the edges, thereby producing less of a pronounced ``dust cloud''. The large fragments behave exactly like in the classical rectangular geometry in figure~\ref{fig:head_comparison}(e) and do not spread alongside the widening shatter plane.
The circular shatter head was selected for its increased spatial spread of the fragments (in all directions), however, comes with a larger uncertainty in the fragment distributions.
The effective shatter angle $\beta$ is a function of the impact position of the pellet inside the tube (defined via the angle $\theta$ indicated in figure~\ref{fig:head_comparison}(d)). Therefore a larger statistical variance was observed for the circular shatter head in the laboratory commissioning phase~\cite{Peherstorfer_2022_fragmentation}. Additionally, the fragments are able to rotate inside the shatter head after impact, which can lead to strong spiral patterns in the resulting fragment plume as shown in figure~\ref{fig:head_comparison}(c). While the motion of the fragments is linear after exiting the shatter head, the spatial distribution -- defined by the angle~$\varphi$ at the exit -- depends on this spiral motion and larger fragments are not only observed close to the upper edge of the exit as illustrated for the rectangular geometry in figure~\ref{fig:head_comparison}(b, e). 
More details on fragment size and velocity distributions are available in the thesis by T.~Peherstorfer~\cite{Peherstorfer_2022_fragmentation}. A fragment segmentation and analysis pipeline based on machine learning is also currently under development~\cite{Illerhaus_2024_SPI}.
Different barrels -- installed ahead of the experiments -- allow to freeze pellets with different diameters. For the 2022 campaign, we installed a set of three 4~mm diameter barrels at first, and during the campaign switched to a set of three 8~mm diameter barrels. The nominal maximum pellet length over diameter~(L/D) of the pellets is 1.7~(4~mm) and 1.2~(8~mm)~\cite{Heinrich_2024_SPI_Lab}. The pellets can be shortened with the help of the barrel heating~(BH) coils.
In the 2022 campaign a total of $\sim$240 dedicated SPI discharges were executed.

\subsection{Radiated power measurements \label{sec:Radiated_power}} 
\begin{table*}[htb]
        \caption{Terms used to calculate the radiated energy fraction~\frad. The parameter reconstruction through function parametrization (FPC/FPG) is described by Braams~\etal~\cite{Braams_1986_FP} and McCarthy~\cite{McCarthy_1992_PhD}.}
        \label{tab:diagnostics_Wth_Wmag}
        \centering
        \begin{tikzpicture}
            \node (table) [inner sep=0pt] {
            \begin{tabular}{l|l|l|l}
                \textbf{term} & \textbf{source / signal} & \textbf{origin / shotfile} & \textbf{short description} \\
                \hline
                \hline
                \Wrad & calculated & eq.~\eqref{eq:Wrad} & total radiated energy from \tFL until \tCQend or \trecovery\\
                \hline
                $\text{P}_\text{rad, sector}$ & calculated & eq.~\eqref{eq:P_rad_sector} & radiated power for one sector\\
                \hline
                $\text{P}_\text{ch}$ & e.g. powF16:0 & BOLZ (BLB) & radiated power for one OV (e.g. channel 0 of sector 16)\\ 
                \hline
                \hline
                \Wpl & calculated & eq.~\eqref{eq:frad_full} & energy dissipated in the plasma~\cite{Lehnen_2013}\\
                \hline
                \hline
                \Wth & \(\text{W}_\text{mhd}\) & AUGD (FPG) & thermal stored energy (including fast particles $\rightarrow$ \(\text{W}_\text{mhd}\))\\
                \hline
                \hline
                \Wmag & calculated & eq.~\eqref{eq:Wmag} & magnetic stored energy\\
                \hline
                $L$ & calculated & eq.~\eqref{eq:L} & plasma inductance\\
                \hline
                $\mu_0$ & constant & $4\pi \cdot 10^{-7}\ \text{N}\text{A}^{-2}$ & vacuum permeability\\
                \hline
                \Ip & IpiFP & AUGD (FPC) & plasma current\\
                \hline
                \(li\) & li &AUGD (FPC) & dimensionless plasma internal inductance\\
                \hline
                \(R\) & Rcurr & AUGD (FPG) & plasma major radius of the current axis\\
                \hline
                \(a\) & ahor & AUGD (FPG) & plasma minor radius\\
                \hline
                \hline
                \Wexheat & calculated & eq.~\eqref{eq:Wexheat} & NBI \& Ohmic heating\\
                \hline
                P(NBI) & PNI & AUGD (NIS) & Neutral Beam Injection (NBI) heating power\\
                \hline
                P(ECRH) & PECRH & AUGD (ECS) & Electron Cyclotron Resonance Heating (ECRH) power\\
                \hline
                P(Ohmic) & calculated & eq.~\eqref{eq:POhmic} & Ohmic heating power\\
                \hline
                \(\text{U}_\text{loop}\) & ULid12f & AUGD (MAU) & loop voltage measurement\\
                \hline
                \hline
                \Wcoupl & estimated & 50\% \Wmag (fig~\ref{fig:Wcoupled_Nina})& magnetic energy coupled into the system (coils, vessel, ...)
            \end{tabular}
            };
            \draw [rounded corners=.5em] (table.north west) rectangle (table.south east);
        \end{tikzpicture}
    \end{table*}

In preparation for the SPI experiments, five new, absolutely calibrated, 4-channel foil bolometers~\cite{Bernert_2013_PhD, Bernert_2014_Bolometer} were installed at five different toroidal locations (sectors) inside AUG. The toroidal positions are (angle to sector of SPI - clockwise): S16 ($0^\circ$), S15 ($22.5^\circ$), S9 ($157.5^\circ$), S5 ($247.5^\circ$ or $-112.5^\circ$), and S1 ($337.5^\circ$ or $-22.5^\circ$) as shown in figure~\ref{fig:exp_setup}(a, c). Hereby, five out of the 16 sectors of AUG are equipped with these foil bolometers sharing the same poloidal geometry, with three of them centered around the injection location in sector 16, one ``perpendicular'' (S5) and one ``opposite'' (S9).

In this manuscript we focus on the analysis of the foil bolometers, which have an effective time resolution of about 0.8~ms, but their measurement directly corresponds to the energy absorbed from the observation volume~(OV)~\cite{Bernert_2013_PhD, Bernert_2014_Bolometer}.
First, the radiated power is calculated for each sector individually:

\begin{equation}
    \text{P}_\text{rad, sector} = \sum_{\text{ch}=0}^{3} w_\text{ch} \cdot \text{P}_\text{ch},
    \label{eq:P_rad_sector}
\end{equation}

with the power measurement \(\text{P}_\text{ch}\) for each individual channel depicted in figure~\ref{fig:exp_setup}(a). The weighting coefficients \(w_\text{ch} = [2.18/16,\ 2.54/16,\ 2.41/16,\ 1.74/16]\) for each of the 4 channels were derived taking into account the geometric effects of the camera. 
The observation volumes for the four channels shown in figure~\ref{fig:exp_setup}(a) (full-shadow volumes) together do not cover the entire volume of the plasma inside each of the five measurement sectors. 
Therefore, weighting factors are introduced to calculate the total radiated power for each of the measurement sectors or even the entire plasma from the measurement in just a single sector. Hereby, a toroidally uniform radiation is assumed within the sector ($\varphi = 22.5^{\circ}$) or the entire plasma ($\varphi = 360^{\circ}$) and  geometrical effects are taken into account -- namely the half/full-shadow (on the sensor) effects of the pinholes to create the channel.
Within this paper, we use the weighting factors divided by 16 (number of sectors in AUG see figure~\ref{fig:exp_setup}(c)) to calculate the radiated power for each individual measurement sector (instead of the entire torus) as given in equation~\eqref{eq:P_rad_sector}. More details on the derivations are available in the PhD thesis by P.~Heinrich~\cite{Heinrich_2025_PhD}.

For the entire calculation methodology described in this section, we assume an uncertainty of about 20\% comparing it to other measurement techniques. However, for a shot-to-shot comparison of the radiated energy values for the SPI experiments, we assume a measurement uncertainty in the range of 10\% as they are determined in the same way and no degradation of the sensitivity of the foil bolometers has been observed over the entire shot range.

\section{Definition and calculation of the radiated energy fraction~\frad \label{sec:frad}} 

The fraction of total plasma energy radiated during the entire disruption -- the radiated energy fraction~(\frad) -- is calculated following the formula by \mbox{Lehnen~\etal~\cite{Lehnen_2013}} and \mbox{Sheikh~\etal~\cite{Sheikh_2020_AUG}}:

\begin{equation}
	\label{eq:frad_full}
	\frad = \frac{\Wrad}{\Wpl} =  \frac{\Wrad}{\Wmag + \Wth + \Wexheat - \Wcoupl},
\end{equation}

with the additional \Wexheat term, representing the plasma heating after the injection. The fixed pre-injection values of~\Wmag (magnetic) and~\Wth (thermal) are used, while~\Wrad (radiated) and~\Wexheat are calculated from the start of the injection until the end-marker, which will lead to~\mbox{$\textrm{f}_\textrm{rad}$(t)} as a function of time, which is evaluated at the end of the integration interval (discussed in the following section~\ref{sec:Wrad}).

\begin{figure}[h]
	\centering
	\includegraphics[width=\linewidth]{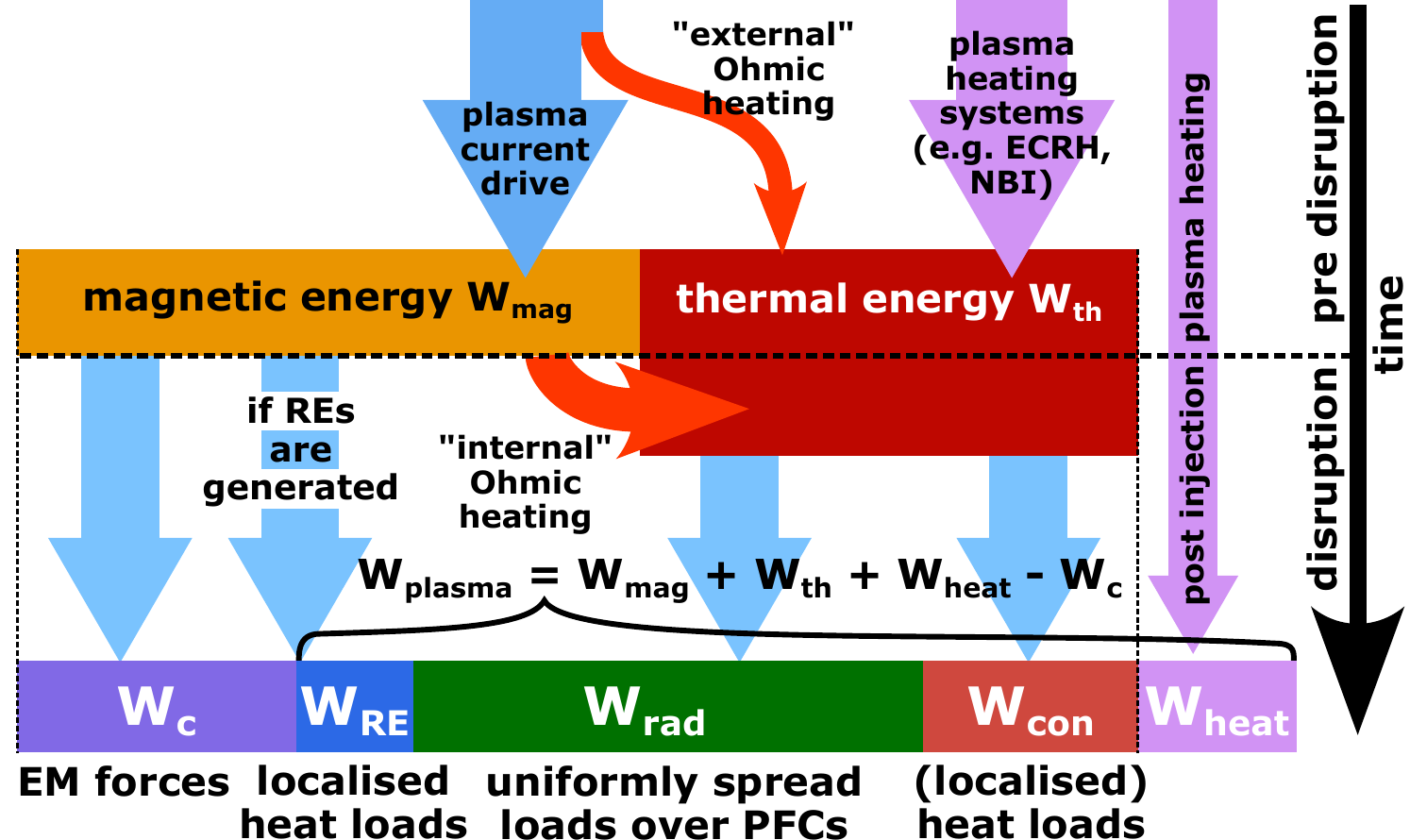}
	\caption{Energy balance during the disruption. The magnetic and thermal energy of the plasma are converted into $\Wcoupl \ (+ \textrm{W}_\textrm{RE}) + \Wrad + \textrm{W}_\textrm{con}$, with $\textrm{W}_\textrm{con}$ being the conducted and convected energy to the plasma facing components. In the AUG SPI experiments, the population of runaway electrons~(REs) is negligible, hence the term $\textrm{W}_\textrm{RE}$ is set to zero for the analysis. The term $\Wexheat$ describes the external plasma heating after the injection (see section~\ref{sec:Wexheat}). The energy dissipated in the plasma~(\Wpl) after the disruption is therefore given as $\Wmag + \Wth + \Wexheat - \Wcoupl$.\label{fig:energy_balance}}
\end{figure}

In the following, the methods to calculate the elements of equation~\eqref{eq:frad_full} are presented.

\subsection{Estimation of the radiated energy~\Wrad} \label{sec:Wrad}
Prior to the first, large fragments entering the plasma, a slight radiation increase in the outermost line of sight~(LOS) (directed towards the low-field side~(LFS) midplane) of the fast AXUV camera \cite{Bernert_2013_PhD, Bernert_2014_Bolometer} in sector 16 is observed, which is referred to as the \textit{\textbf{F}irst \textbf{L}ight}~(FL), and is used as the marker for the beginning of the injection phase. This is assumed to be caused by the fragments or small amounts of gas (produced during the pellet break-up) entering the plasma followed by the bulk of the fragments shortly afterwards. The term \(\text{W}_\text{rad}\) is the estimate for the energy radiated from the entire plasma measured via the foil bolometers, where linear interpolations of sectors with no measurements are used. As a start of the integration interval the FL-marker~(\tFL) is used. The end-marker~(\tend) is provided by the end of the radiation peak(s), when the derivative of the radiation approaches zero again. The methodology used to derive this end-marker is described in the paper by P.~Heinrich~\etal~\cite{Heinrich_2024_CQmodes} where it is applied to the plasma current signal to detect the end of the CQ. For disruptive SPI, this end-marker is usually at the end of the CQ~(\tCQend). For a non-disruptive injection, the end-marker is usually around the minimum of the plasma current, before the plasma current starts to recover to its pre-injection level.
    
The radiated power is calculated for each individual sector, where for sectors without direct measurements, a linear interpolation between the measurement sectors was chosen.
Here, the highest peaking is observed for low neon or 100\%~$\textrm{D}_2$ injections in sector~16 (injection location), where the neighboring sectors~1 and~15 already show radiation levels very close to the sectors~5 and~9, hence, a linear interpolation was suggested~\cite{Heinrich_2023_EPS}.
Finally, the total radiated power is given by the sum over all 16 sectors:
    
    \begin{align}
    	\label{eq:Wrad}
	\Wrad &= \int_{\tFL}^{\tend} \text{P}_\text{rad, total}\ \dt\\
	&= \int_{\tFL}^{\tend} \sum_{\text{sector}=1}^{16} \text{P}_\text{rad, sector}\ \dt, \notag
    \end{align}
    
with $\text{P}_\text{rad, sector}$ as defined in equation~\eqref{eq:P_rad_sector} and afterwards integrated over time to obtain \Wrad.

\subsection{Plasma stored energies}
The two terms \Wth and \Wmag describe the respective thermal and magnetic stored energies prior to the FL. The stored magnetic energy is calculated as~\cite{Loarte_2011_Magnetic, Igochine_2015_MHD}
    
    \begin{equation}\label{eq:Wmag}
        \Wmag = 0.5 \cdot L \cdot \Ip^2,
    \end{equation}
    
with the plasma inductance \(L\)~\cite{Lehnen_2013, Loarte_2011_Magnetic}
    
    \begin{equation}\label{eq:L}
        L = \mu_0 \cdot R \cdot \left[0.5\cdot\text{li} + \text{ln}(8R/a) - 2\right].
    \end{equation}

The different terms and their reconstruction/signal names are given in table~\ref{tab:diagnostics_Wth_Wmag}. Hereby, the pre-injection values of \Wth and \Wmag are given as the average of the signal from 50~ms before \tFL until \tFL.

\subsection{Additional heating power~\Wexheat \label{sec:Wexheat}}
The term \Wexheat is added to the calculation as the heating systems typically do not shut down instantly at the time of the FL, but continue heating the plasma (compare plasma heating after \tFL in figure~\ref{fig:scenario_plots}(b)). For this calculation it is assumed, that all the energy of the heat sources is absorbed. Due to this assumption, the resulting values of \frad in figure~\ref{fig:Wrad_vs_Wplasma}(b) can be interpreted as a lower limit compared to the definition of~\Wpl without \Wexheat, which may serve as an upper limit. Automatic shut-down of external heating systems typically occurs when each system detects unsatisfactory absorption (i.e. NBI shine-through or ECRH reflection). Therefore, \Wexheat is calculated as
    
    \begin{equation}\label{eq:Wexheat}
        \Wexheat = \int_{\tFL}^{\tend} P_\text{NBI} + P_\text{ECRH} + P_\text{external Ohmic},
    \end{equation}
    
with the heating via the Neutral Beam Injection~(NBI), Electron Cyclotron Resonance Heating~(ECRH) and Ohmic heating. The Ohmic power~$P_\text{Ohmic}$ is estimated as
    
    \begin{equation}\label{eq:POhmic}
        P_\text{external Ohmic} = \Ip \cdot U_\text{loop}.
    \end{equation}

The $U_\text{loop}$ measurement at the plasma edge is meant to be representative of the external Ohmic power input from the central solenoid during the disruption. On the time scales considered, the electric field induced in the core typically does not propagate to the edge measurement~\cite{Papp2011runaway,Papp2013effect}.

\subsection{Coupled energy~\Wcoupl \label{sec:Wcoupl}}
\begin{figure}[h]
	\centering
	\includegraphics[width=\linewidth]{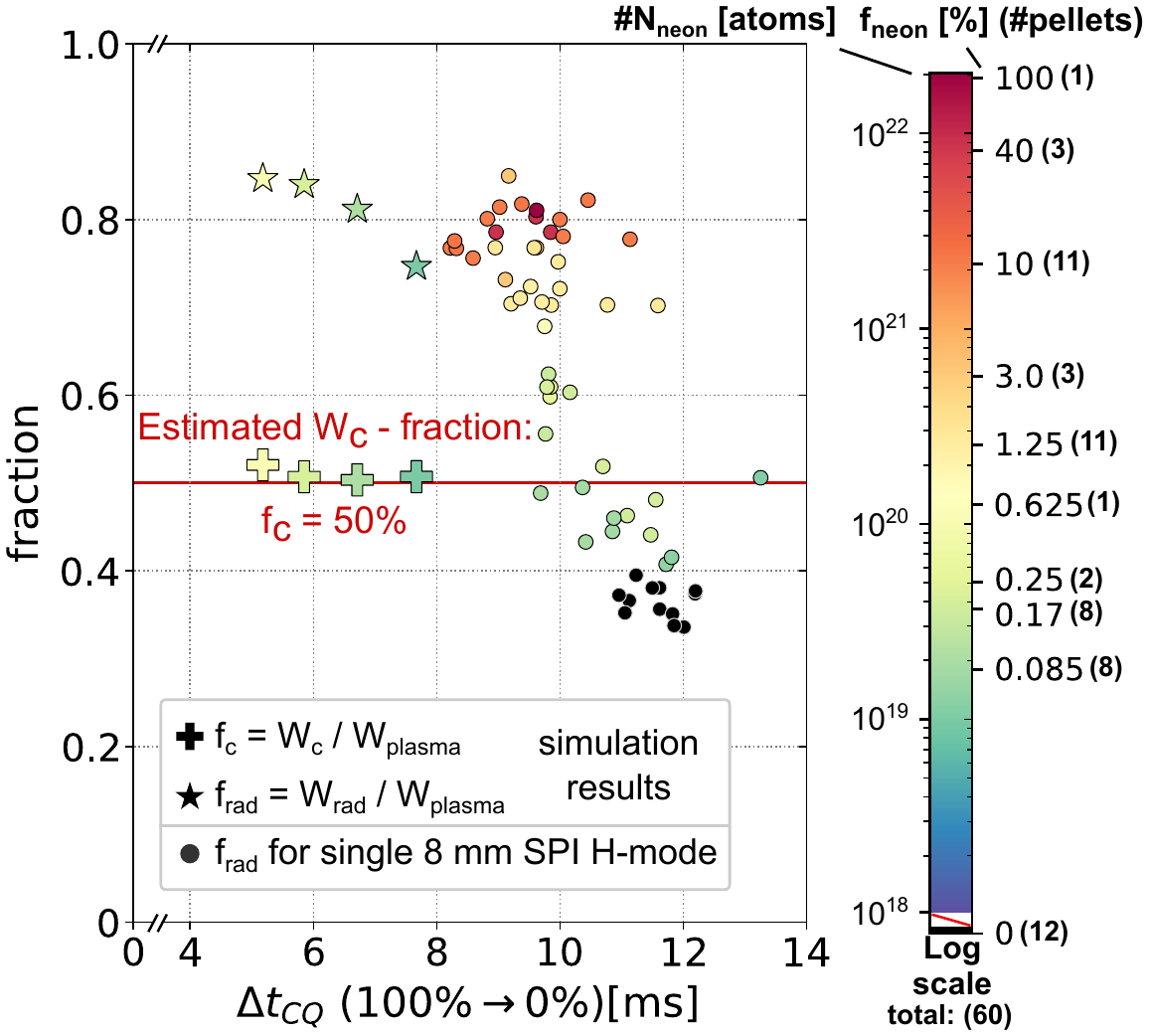}
	\caption{Radiated energy fraction (\frad) and coupled energy fraction (\fc) plotted against the current quench duration ($\Delta t_\text{CQ}$) from the IP-spike (100\%) to the end of the CQ (0\%). Different symbols indicate the coupled fraction~(+) and the radiated energy fraction~(stars) from the simulations as well as the experimental \frad-values~(circles) for 8~mm diameter pellets into the SPI H-mode. The colour-coding indicates the number of neon atoms inside the pellet for the experiment (estimated) and simulation. On the right of the colourbar, the neon content in percent (derived from partial pressure in the mixing volume) and the number of pellets in the brackets is provided. To calculate the experimental \frad-values (compare circular points and figure~\ref{fig:Wrad_and_frad}(b)), a ratio of \Wcoupl to \Wmag of 50\% is assumed.\label{fig:Wcoupled_Nina}}
\end{figure}

As there is no direct measurement of the entire coupled energy \Wcoupl into the surrounding structure, \Wcoupl is estimated as 50\% of the pre-injection magnetic energy, based on the observations by \mbox{Sheikh~\etal~\cite{Sheikh_2020_AUG}} and JOREK simulations for ASDEX Upgrade VDEs performed by Schwarz~\cite{Schwarz_2024_PhD} shown in figure~\ref{fig:Wcoupled_Nina}. (We note, that, coincidentally, a value of 50\% was also estimated for JET by \mbox{Lehnen~\etal~\cite{Lehnen_2013}}.)
In order to estimate the coupled energy, 2D simulations with the extended MHD code JOREK~\cite{Hoelzl_2021} were carried out based on an L-mode discharge~\cite{Schwarz_2023}.
Different quantities of neon were introduced with a uniform source in a post-TQ plasma, leading to different CQ durations.
One can calculate the radiated energy and the dissipated energy by Ohmic heating $\text{W}_\text{internal Ohmic}$ during the CQ phase.
As the magnetic energy can either be lost by internal Ohmic heating or by coupling to the magnetic structures (and the
RE population is negligible), the coupled energy is estimated as:
\begin{equation}
	\Wcoupl = \Wmag - \text{W}_\text{internal Ohmic}\label{eq:Wcoupl}.
\end{equation}

The results are shown in figure~\ref{fig:Wcoupled_Nina}, where the fraction of the magnetic energy that is coupled to the conducting structures (\fc) and the radiation fraction of the available magnetic energy (\frad) against $\Delta \text{t}_{CQ}$ are shown. 
Note, that in the simulation the external heating sources (NBI, ECRH) are stopped at the start of the disruption.
As the neon source is added after the TQ, these quantities only refer to the CQ phase.
These simulations can be used as a basis to validate a lumped parameter model of the conductive structures to estimate the coupled energy as done by \mbox{Lehnen~\etal~\cite{Lehnen_2013}}.
For low neon content (\mbox{$1-5 \times 10^{18}$} atoms) the disruptions are no longer radiation dominated, a direct comparison with the experimental results becomes increasingly difficult, and beyond the scope of this simple \Wcoupl estimate.
The duration of the CQ is calculated between the IP-spike~(100\%) and the end of the CQ~(0\%)~\cite{Heinrich_2024_CQmodes} for the experiments and simulations for better comparability.
A wall time of about 60~ms is used in these simulations~\cite{Schwarz_2024_PhD}. We expect the coupled energy~\Wcoupl to depend strongly on the wall time for the net vessel current. The shorter this time is, the less energy is coupled into the vessel. At the time these simulations were performed, the complicated wall structures were not well represented in the coupled JOREK-STARWALL simulations. Using the toroidal resistivity of $3\ \mu \Omega\textrm{m}$ derived by \mbox{Giannone~\etal~\cite{Giannone_2015}} instead of the $0.45\ \mu \Omega\textrm{m}$ used in the simulations, we expect the wall time to be in the range of $\sim 9$~ms instead. While for short CQ durations ($< 10$~ms) the 50\% coupled energy remains a good approximation, the coupled energy may be overestimated for longer CQ durations as observed for the low neon and 100\%~$\textrm{D}_2$ SPI. Ultimately, the radiated energy fraction of these discharges from section~\ref{sec:Wrad_frad_values} might be lower.

\subsection{Plasma scenarios of the 2022 SPI experiments \label{sec:Scenarios}}

\begin{figure*}[htb]
	\centering
	\includegraphics[width=\linewidth]{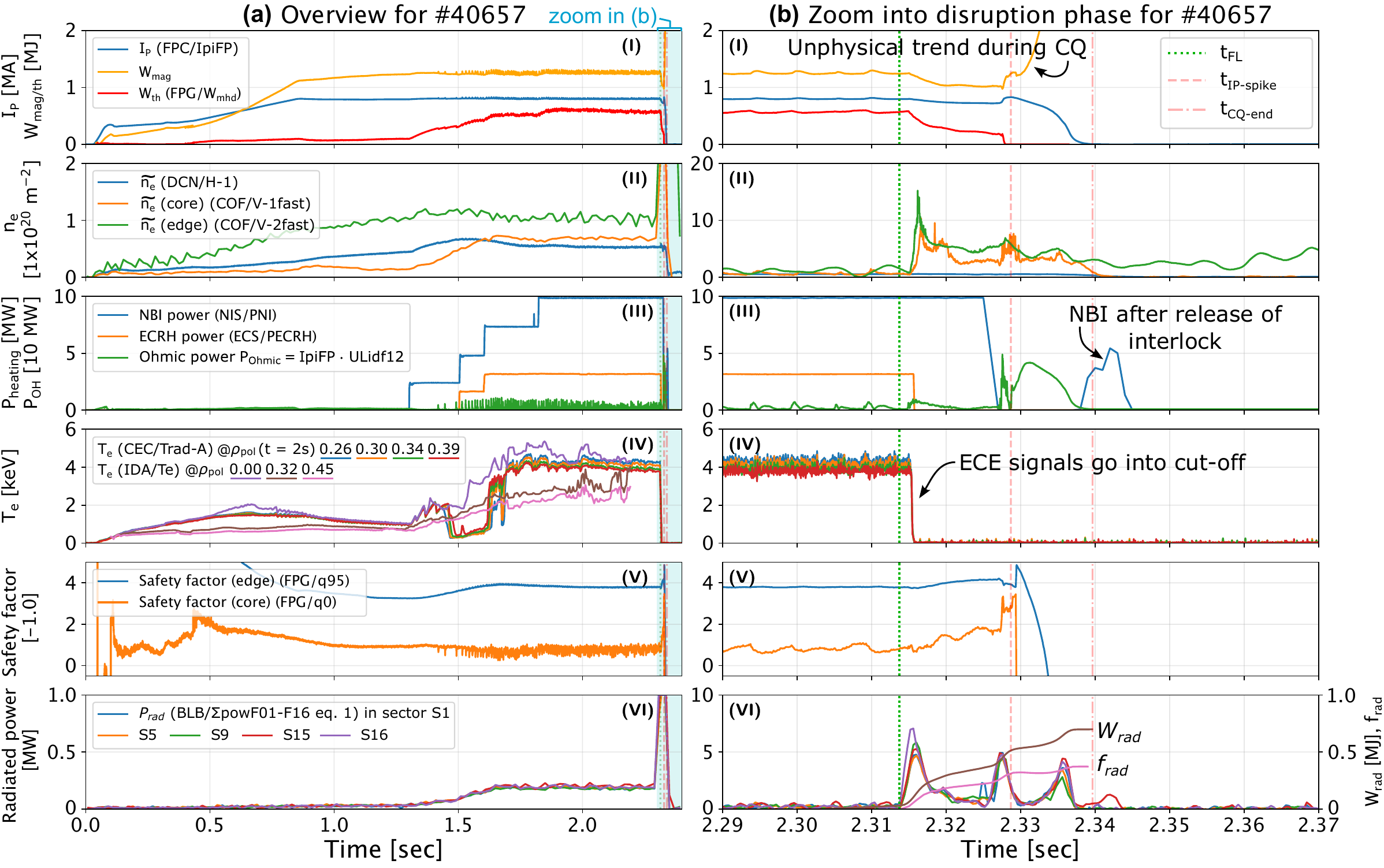}
	\caption{Overview plot for the ``SPI H-mode'' scenario from table~\ref{tab:Scenarios}. In~(a) the entire discharge is provided, while in~(b) the injection and disruption part is shown in more detail. The disruption is initiated by injection of a 100\%~$\textrm{D}_2$ injection, triggered at 2.30~s. Note the different y-axis limits between~(a) and~(b) in~(II) and~(VI). The following signals are provided:
	In~(I), plasma current, stored magnetic and thermal energies are provided. The line integrated ($\textrm{m}^{-2}$) electron densities derived from the core DCN~\cite{Mlynek_2010_DCN, Mlynek_2014_Fringe-jumps} and fast CO$_2$ lasers~\cite{Mlynek_2012_CO2} are displayed in~(II). Hereby, the fast CO$_2$ laser signals are smoothed via a rolling window average with window size of $\textrm{w}_\textrm{avg} = 16$~ms in~(a). In~(III), the heating powers (NBI, ECRH, Ohmic) are shown. In~(IV), the electron temperature via the electron cyclotron emission~(ECE)~\cite{Denk_2018_AUG-ECE} measurement for different radial channels, with $\textrm{w}_\textrm{avg} = 3.2$~ms in~(a) as well as the $\textrm{T}_\textrm{e}$ reconstruction from the integrated data analysis~(IDA)~\cite{Fischer_2010_IDA}, with $\textrm{w}_\textrm{avg} = 5$~ms in~(a). Note, that the ECE signals go into cut-off at $\approx 2.315$~sec, hence do not reflect the TQ time (compare \Wth in~(I)). In~(V), the core and edge (at \mbox{$\rho_\textrm{pol} = 0.95$}) safety factors are given. In~(VI), the total radiated power in the five toroidal sectors derived from the foil bolometers (equation~\eqref{eq:P_rad_sector}) is provided, with $\textrm{w}_\textrm{avg} = 20$~ms in~(a). In~(VIb), the total radiated energy (equation~\eqref{eq:Wrad}) and radiated energy fraction (equation~\eqref{eq:frad_full}) are provided. In this case, three large radiation peaks are visible at: fragment arrival, the TQ/IP-spike, and the final VDE phase.}
	\label{fig:scenario_plots}
\end{figure*}

In this section, the different plasma scenarios for the 2022 SPI experiments are introduced. 
An overview plot for the typical ``SPI H-mode'' scenario is provided in figure~\ref{fig:scenario_plots} and the main plasma parameters for the different scenarios are summarised in table~\ref{tab:Scenarios}.
Note, that the ECE signals displayed in figure~\ref{fig:scenario_plots}(IV) go into density cut-off, hence do not represent the thermal quench~(TQ) or global reconnection event~(GRE) time as indicated by the \Wth signal in figure~\ref{fig:scenario_plots}(I).
The pellet(s) are usually injected around 2.3~seconds into the flat top phase of the discharge, with minor intentional changes in timing due to various reasons (e.g. trying to match pellet arrival with Thomson lasers). The heating power in table~\ref{tab:Scenarios} refers to the approximate heating power until the shut down of the heating systems during the disruption. 
The thermal fraction~\fth~\cite{Lehnen_2013, Sheikh_2020_AUG} is defined as

\begin{equation}
	\label{eq:fth}
	\fth = \frac{\Wth}{\Wpl(\mathrm{t}_\mathrm{FL})} =  \frac{\Wth}{\Wmag + \Wth - \Wcoupl}.
\end{equation}

\begin{table*}[htb]
        \caption{Plasma scenarios for SPI experiments in 2022 (see also figure~\ref{fig:Wrad_vs_Wplasma}). The plasma parameters are averaged over 50~ms prior to the \tFL. The line integrated electron density is taken from the H1 (DCN) LOS. The average thermal fraction~\fth is calculated via equation~\eqref{eq:fth}. The plasma current was around 800~kA besides the ``High~\Wpl'' discharges at 1~MA.}
        \label{tab:Scenarios}
        \centering
        \begin{tikzpicture}
            \node (table) [inner sep=0pt] {
            \begin{tabular}{c|c|c|c|c|c|c|c}
                \textbf{plasma} & \textbf{$\text{B}_\text{tor}$} & \multicolumn{2}{c|}{\textbf{$\text{P}_\text{heating}$ [MW]}} & \textbf{$\text{q}_\text{95}$} & \textbf{line integr. electr.} & \textbf{avg. pre-FL} & \textbf{avg.}\\
                 \textbf{scenario} & \textbf{[T]} & $\text{P}_\text{NBI}$ & $\text{P}_\text{ECRH}$ & & \textbf{density $\text{n}_\text{e}$ [$\times 10^{19}\ \text{m}^{-2}$]} & \textbf{\Wpl [MJ]} & \textbf{\fth}\\
                \hline
                \hline
                SPI H-mode & 1.8 & 7.5--9.9 & 2.85--3.2 & $< 4$ & 4.5--6.8 & 1.25--1.4 & 0.48\\
                \hline
                2.5T H-mode & 2.5 & 9.3--9.9 & 2.85--3.0 & $> 5$ & 5.5--6.8 & 1.3--1.4 & 0.48\\
                \hline
                Low Te & 1.8 & 2.4 & 1.6 & $< 4$ & 5.5--6.1 & 0.93 & 0.33\\
                \hline
                High \Wpl & 2.5 & 8.3--9.7 & 2.05 & $\approx 4$ & no measurement & 1.92 & 0.47\\
                \hline
                Ohmic & 1.8 & -- & -- & $< 4$ & 3.9--4.5 & 0.69--0.74 & 0.15 
            \end{tabular}
            };
            \draw [rounded corners=.5em] (table.north west) rectangle (table.south east);
        \end{tikzpicture}
\end{table*}

\section{Experimental values of \Wrad and \frad \label{sec:Wrad_frad_values}}

\begin{figure*}[htb]
	\centering
	\includegraphics[width=\linewidth]{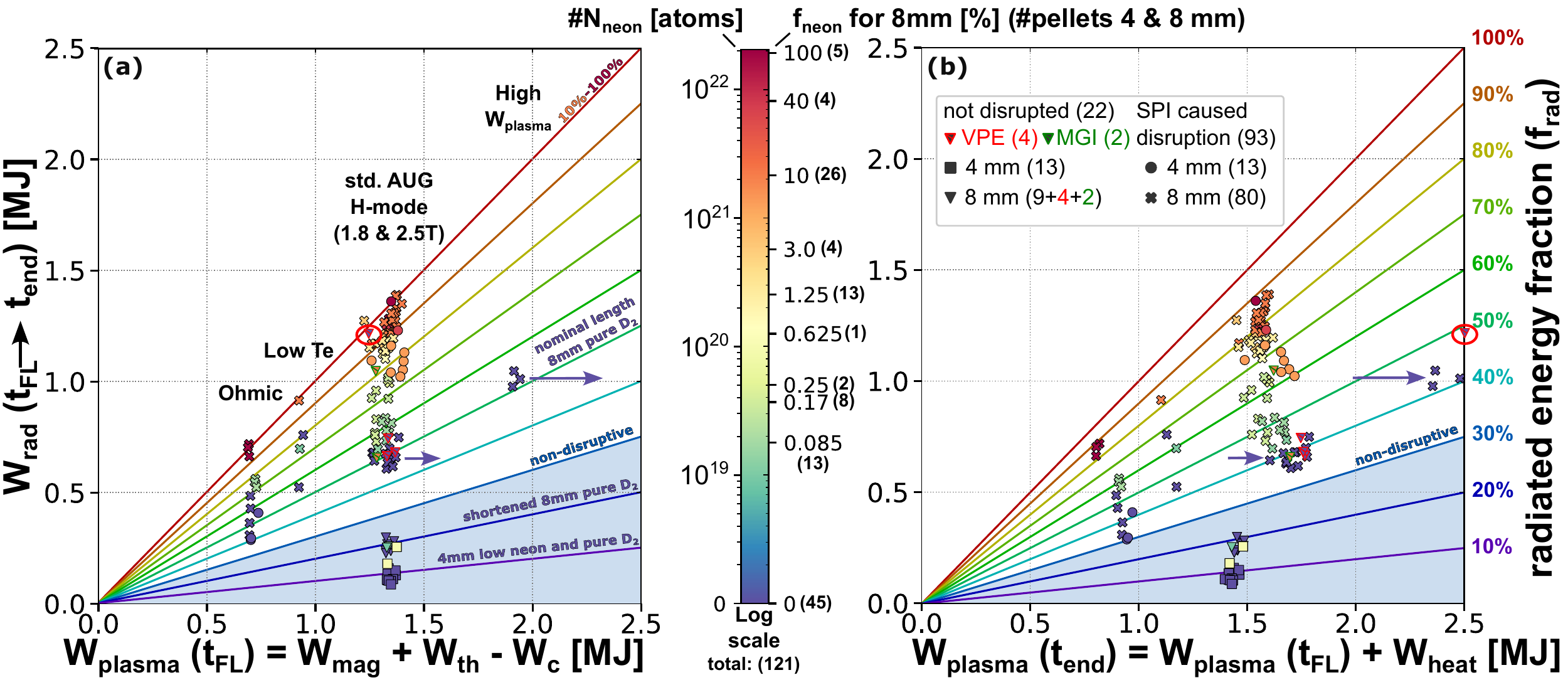}
	\caption{The radiated energy during SPI, plotted against the energy dissipated in the plasma~\Wpl~\cite{Lehnen_2013} (see equation~\eqref{eq:frad_full}) at \tFL and \tend, respectively. The five different target plasmas indicated in~(a) are introduced in table~\ref{tab:Scenarios}. The colour of the points represents the amount of neon inside the pellets as indicated by the colourbar in the center. The estimated number of neon atoms (left colour-scale) which sets the point colour and the equivalent partial pressure ratio in the mix tank for the 8~mm pellets (right). The number inside the bracket represents the number of pellets (including 4 and 8~mm pellets). The \Wexheat-term in~(b) shifts \Wpl to higher values (to the right) indicated by the arrows and red circled value, depending on pre-TQ duration. Therefore, an upper limit on the \frad-values is provided in~(a) and a lower limit in~(b) as this assumes 100\% absorbtion of \Wexheat.}
	\label{fig:Wrad_vs_Wplasma}
\end{figure*}

\begin{figure*}[htb]
	\centering
	\includegraphics[width=\linewidth]{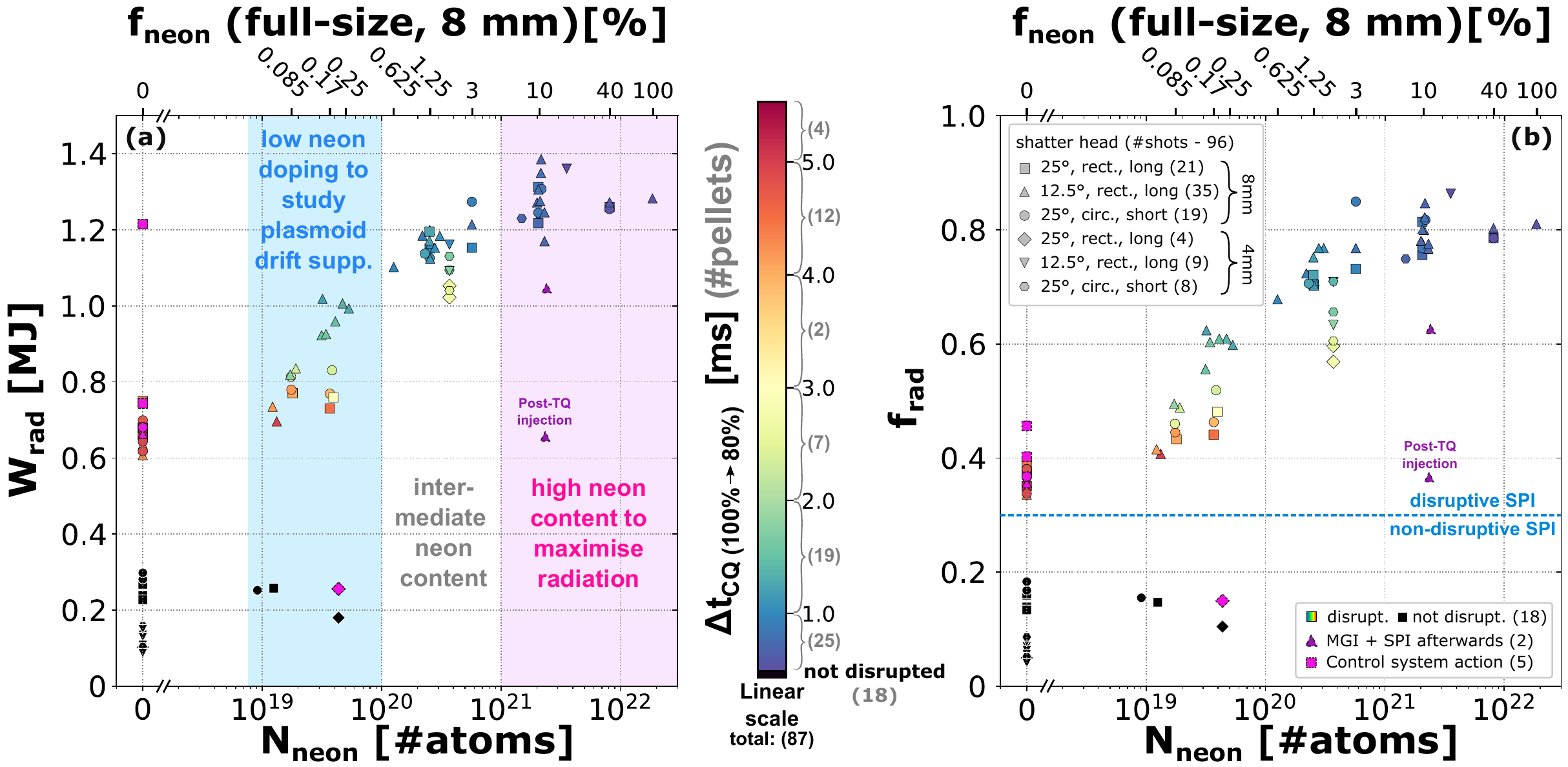}
	\caption{(a) Radiated energy and (b) radiated energy fraction \frad as a function of the estimated number of neon atoms inside the pellet. The different point types represent the different shatter heads for the 4 and 8~mm diameter pellets. The colour bar shows the length of the CQ time (100\% $\rightarrow$ 80\% with 100\% being the IP-spike level) in a linear scale. Black points are non-disruptive shattered pellet injections. The two purple points show SPI into massive gas injection~(MGI) triggered disruptions. Magenta points are SPI injections in which an action in the discharge control system~(\DCSVPEState{}) was triggered by the injection, hence might affect the disruption dynamics (more detail in the text).}
	\label{fig:Wrad_and_frad}
\end{figure*}

In this section, the \Wrad and resulting \frad values for the 2022 experimental campaign at ASDEX Upgrade are presented.
Figure~\ref{fig:Wrad_vs_Wplasma} shows the total radiated energy~\Wrad plotted against the energy dissipated in the plasma~\Wpl as defined in equation~\eqref{eq:frad_full}. The colour-coding shows the neon content inside the pellet ranging from 100\%~$\textrm{D}_2$ to 100\%~Ne pellets on a logarithmic scale, markers represent the 4~mm and 8~mm pellets for (non-)disruptive injections. Numbers in the brackets indicate the number of pellets in that category. Note, that while the neon partial pressure ratio ($\textrm{f}_\textrm{neon}$ -- right side of the colourbar) only corresponds to the 8~mm pellets, the number in the brackets indicate the total number of 4 and 8~mm pellets with this neon percentage. This partial pressure ratio~($\textrm{f}_\textrm{neon}$) for the 8~mm pellets is used for the 2nd x-axis in figures~\ref{fig:Wrad_and_frad} and~\ref{fig:shatter_head_fits}. The target plasma scenarios from section~\ref{sec:Scenarios} are indicated in figure~\ref{fig:Wrad_vs_Wplasma}(a).

In figure~\ref{fig:Wrad_vs_Wplasma}(a) the plasma stored energy is evaluated for the time of the first light, hence without the additional~\Wexheat term, consequently the scenarios occur at distinct values of~\Wpl. The H-mode scenarios at 1.8~T (SPI~H-mode) and 2.5~T share the same~\Wpl.
As previously reported for \mbox{DIII-D}~\cite{Shiraki_2016}, increasing the amount of neon inside the pellet leads to an increase in radiated energy fraction from around 10\% (4~mm diameter 100\% $\textrm{D}_2$ pellets) to 100\% (high neon concentrations of 10\% and beyond for the 8~mm diameter pellets). Without additional heating, the energy dissipated in the plasma given by~\Wpl~\cite{Lehnen_2013}. The total dissipated energy including external heating is summarised as the term~\Wpl(\tend) (figure~\ref{fig:Wrad_vs_Wplasma}(b)). Hereby, long disruption phases (lower neon content) are shifted further to the right, as typically more heat is injected into the plasma until the heating systems are switched off. The radiated energy fraction of non-disruptive SPI is below 30\% (indicated by the blue area in figure~\ref{fig:Wrad_vs_Wplasma}).

Figure~\ref{fig:Wrad_and_frad}(a) shows the radiated energy and figure~\ref{fig:Wrad_and_frad}(b) the radiated energy fraction~(\Wrad/$\text{W}_\text{plasma}(\tend)$ compare right axis in figure~\ref{fig:Wrad_vs_Wplasma}) as a function of the neon content for single pellet SPI. The colour-coding (linear scale) represents the measured CQ duration from 100\% (peak value of the IP-spike) to 80\% of the plasma current, with black points representing the non-disruptive injections. For long disruption durations (low neon content) often a vertical displacement event~(VDE) is triggered at some point in time. For a better inter-shot comparison, the 80\% marker is used for the early CQ duration~\cite{Jachmich_2023_EPS} prior to the VDE dynamics.

With increasing neon content inside the pellet, the radiated energy fraction also increases.
However, \frad saturates around 80\% when we take the \Wexheat into account. Whether this energy is not absorbed by the plasma in the first place, conducted to the PFCs as heat, or if potential other loss mechanisms are at play, is presently not identified.
Note, that the coupled energy~\Wcoupl for the low neon content/100\%~$\textrm{D}_2$ with long CQ~times might be overestimated from the present simulations (see section~\ref{sec:Wcoupl}), hence the actual  radiated energy fractions for these injections may be lower.
For disruptive SPI, the radiated energy fraction is between 30\% (100\%~$\textrm{D}_2$ SPI) and 80--85\% (\mbox{$>10^{21}$~neon} atoms).
This is in line with previous results from JET SPI, where most radiated energy fractions were between 30\% and 85\%~\cite{Sheikh_2021}. However, assessing the radiated energy fraction in JET is challenging given the bolometric setup, hence either toroidally symmetric radiation is assumed~\cite{Sheikh_2021}, different toroidal mode-locking positions are examined~\cite{Jachmich_2022_JET-SPI} or 3D simulation codes like Emis3D are applied~\cite{Sheikh_2021, Stein-Lubrano_2024_Emis3D}.
For \mbox{DIII-D}, radiated fractions of $\Wrad/\Wth \lessapprox 90\%$ are reported~\cite{Shiraki_2016}.

For two cases, a disruption was triggered by 100\%~$\textrm{D}_2$ massive gas injection~(MGI) and an 8~mm SPI pellet with 10\% neon was injected into the disrupting plasma (purple triangles). In the first realisation, the fragments of the SPI arrived after the thermal collapse during the thermal quench~(TQ), hence the radiated energy (and radiated energy fraction) is low. With reduced time delay between MGI and SPI, a higher \Wrad (\frad) value was achieved.

\begin{figure*}[htb]
	\centering
	\includegraphics[width=\linewidth]{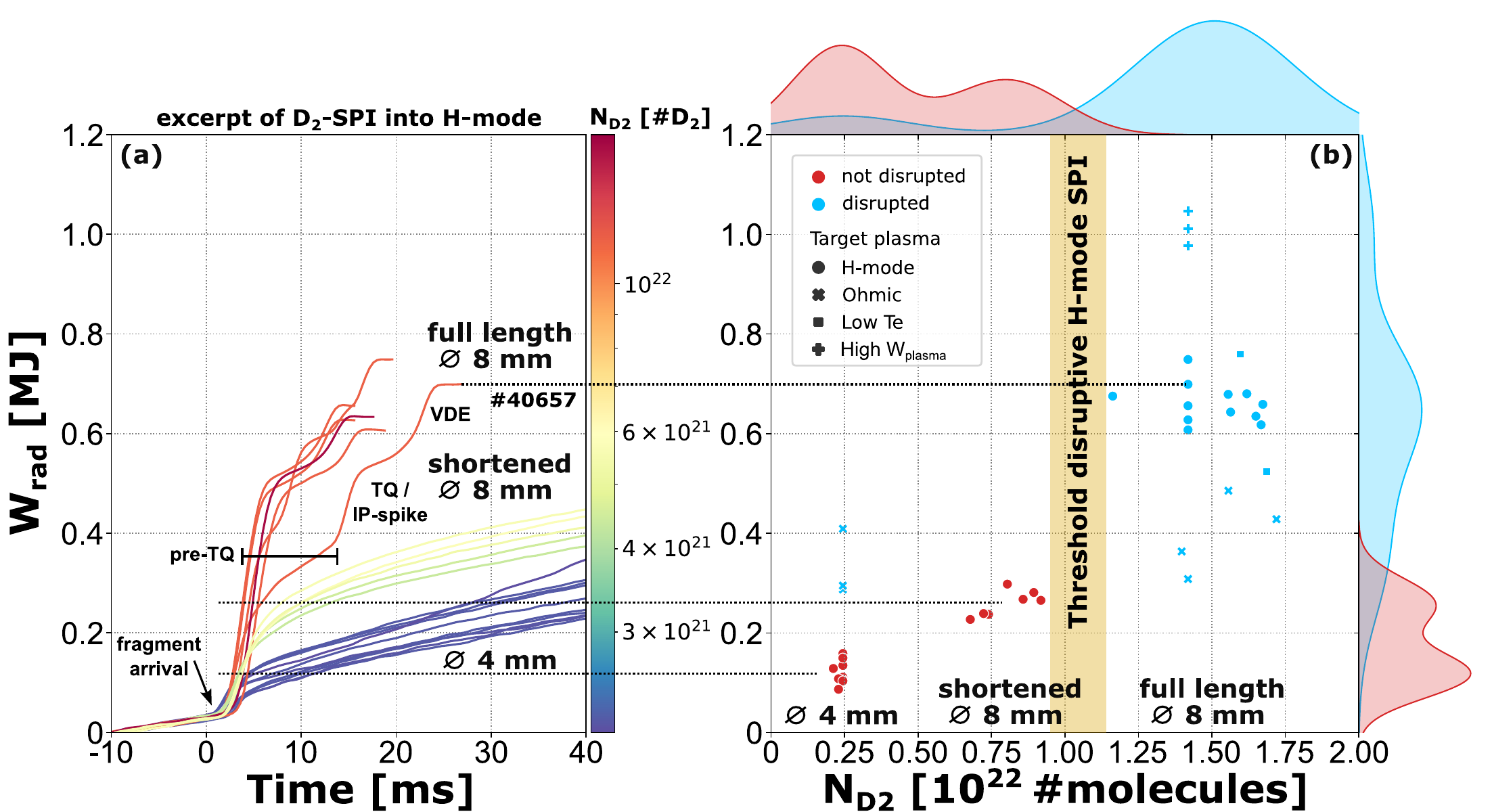}
	\caption{(a) Radiated energy as a function of time and (b) estimated number of deuterium molecules~($\text{D}_2$) for 100\%~$\textrm{D}_2$ SPI into the \mbox{SPI~H-mode}. Only discharges are displayed, in which the \VPEState{}-change of the DCS happened at or after the IP-spike. In~(a), a subset of the H-mode injections is shown with the number of deuterium molecules as the colour-code. The time $t = 0$~ms is with respect to the start of the integration interval (before the first radiation peak) which is around \tFL. The radiation increase for $t < 0$ indicates the pre-injection radiation levels, hence is similar to the radiated power after the recovery for non-disruptive SPI (compare slopes). Three branches emerge with the (full length) 4~mm, 8~mm shortened pellets and the full length 8~mm pellet injections. The threshold of disruptive injections is around \mbox{$1\times 10^{22}$} deuterium molecules in line with previous simulation results~\cite{Hoelzl_2020_D2-SPI}. Note, this disruption threshold also depends strongly on the actual material assimilation, hence the fragmentation \& ablation properties for this discharge, with the number of injected particles used as an approximation. The discharge \#40657 exhibits a long pre-TQ phase -- compare other disruptive cases in red-orange in~(a) -- and might be close to the threshold of being non-disruptive (yellow-green lines).}
	\label{fig:Wrad_N_D2}
\end{figure*}

The magenta points in figure~\ref{fig:Wrad_and_frad} indicate discharges in which a discharge control system~(DCS)~\cite{Sieglin_2020_DCS, Kudlacek_2024_DCS, Weiland_2024_DCS} action was triggered by the SPI (\DCSVPEState{} changed from state 0 to a different state) during the pre-TQ phase. If the DCS was unable to recover the discharge~\cite{Sieglin_2020_DCS} and detects a plasma state that could lead to a disruption (e.g. the plasma current center position passes a $z_\textrm{current}$-position threshold to detect VDEs) it goes into a ``holding'' state (indicated by the state-change of the \VPEState{}), before trying to ramp down the discharge safely if the plasma does not disrupt before the DCS can take any action. Hereby, in most cases the threshold of the allowed vertical displacement of the plasma was exceeded which triggered the state-change of the DCS. The resulting action of the control system might have affected the development of some of these disruptions and in some cases we understand, that the attempt to ramp down the plasma even caused the disruption for injections which -- if the threshold value of the \VPEState{} would have been set to higher values -- might be recoverable via the DCS: We observe that some discharges (either due to interlocks in the NBI or the \VPEState-change and consequent holding-action) do disrupt which otherwise were already about to recover where the $z_\textrm{current}$ position was slowly going back to pre-injection levels and below the trigger value of the \VPEState{}. One of the implications of the holding-state is, that only three NBI sources are allowed to soft-land the discharge (ramp-down), while four were requested for our scenario, i.e. in these cases one NBI source is switched off. However, without the desired beam power for the specific scenario during the holding-state, this might be actually triggering the disruption. Eventually, at some point inside the CQ the \VPEState{}-change may be triggered, however, only discharges where the \VPEState{} changed before the IP-spike are selected for the analysis in this work and excluded from figures~\ref{fig:Wcoupled_Nina},~\ref{fig:Wrad_vs_Wplasma},~\ref{fig:Wrad_N_D2},~and~\ref{fig:shatter_head_fits}.
However, it was observed in some cases (potentially all/a large fraction of cases are affected) that all NBI sources are switched-off shortly before the \VPEState{}-change of the DCS is triggered due to interlocks in the NBI system (e.g. to prevent excessive shine-through). As a consequence a more detailed study on the impact of the switch-off of the heating systems on the pre-TQ time would be required.

The 100\%~$\textrm{D}_2$, 4~mm diameter (or even a few low neon doped) pellets did typically not cause the SPI~H-mode plasmas to disrupt. While the full-sized, 8~mm, 100\%~$\textrm{D}_2$ pellets caused a disruption, the shortened (approx. half nominal pellet length~\cite{Heinrich_2024_SPI_Lab}) pellets did not. For an excerpt of 100\%~$\textrm{D}_2$ SPI in figure~\ref{fig:Wrad_N_D2}, the time evolution of the radiated energy with the colour-coding representing the estimated number of deuterium molecules is shown in~(a). In figure~\ref{fig:Wrad_N_D2}(b) the radiated energy at the end of the integration interval is plotted as a function of estimated number of deuterium molecules. Note, that for some pellets (especially the 4~mm) no measurement of the pellet length was available via the OPD camera. For these pellets, the estimated number of deuterium molecules is taken from pellets with the same recipe for desublimation and launching from the AUG SPI database~\cite{Heinrich_2024_SPI_Lab} which results in the same number of estimated deuterium molecules (vertically aligned points). Estimation from the gas reservoir pressure drop alone is difficult due to the various heatings applied~\cite{Heinrich_2024_SPI_Lab}.
Only discharges which had their \VPEState{} change of the DCS after the pre-TQ phase (during the IP-spike time window or CQ phase) are included in figure~\ref{fig:Wrad_N_D2}.
For the discharge \#40657 ($25^{\circ}$, circular shatter head; \mbox{$\sim 460$~m/s} pre-shatter velocity; see SPI H-mode overview figure~\ref{fig:scenario_plots}) we do not have a direct measurement of the pellet length, hence the value of around \mbox{$1.4\times10^{21}$} deuterium molecules is inferred from the SPI recipe database. However, this discharge exhibits a long pre-TQ phase (around 10~ms) and also lower initial radiation during the pre-TQ phase compared to the other injections depicted in figure~\ref{fig:Wrad_N_D2}(a). Consequently, we suspect this discharge to be close to being non-disruptive. The \VPEState{} change happened for this discharge right at the IP-spike (which is very early compared to other discharges displayed in figure~\ref{fig:Wrad_N_D2} where at some point during the CQ the \VPEState{}-change is triggered), hence the discharge would probably be terminated by the loss of one of the NBI sources due to the control system otherwise. The observation of a disruption threshold around \mbox{$1\times 10^{22}$} deuterium molecules is well in line with simulation results by \mbox{H\"olzl~\etal~\cite{Hoelzl_2020_D2-SPI}}. Scanning the amount of deuterium inside the pellets, the simulation results suggest incomplete TQs below a threshold of \mbox{$(8\pm 4)\times 10^{21}$} deuterium atoms. Note, that the difference of an factor of two between the simulations and experiments might be also caused by the simulation setup: In the simulations, all of the material was assumed to be injected in the form of solid fragments, while in the experiment, part of the pellet material will be transformed into gas during the shattering process which might not (quickly) penetrate deeply into the plasma. Also the fragment size and velocity distribution will impact the amount of assimilated $\textrm{D}_2$, hence the disruption threshold. It depends on the assimilated pellet material, however, only indicated by the injected amount in figure~\ref{fig:Wrad_N_D2}(b).
Compared to the threshold of the experiments for H-modes, we observed Ohmic discharges to disrupt already for the 4~mm diameter pellets (see figure~\ref{fig:Wrad_N_D2}(b)).

While the non-disruptive SPI injections have a radiated energy fraction below 30\%, the fraction increases during the TQ, CQ, and VDE phase of disruptive injections to values of \mbox{$\leq 40$\%} in total.
As indicated by discharge \#40657 in figure~\ref{fig:Wrad_N_D2}(a), the radiation of the full-sized, 8~mm diameter pellet in the pre-TQ phase is only slightly higher compared to the non-disruptive injections into H-mode plasmas.  
When adding the radiated energy of the TQ and also the VDE phase on top of the radiation at the fragment arrival, the difference in \frad between the non-disruptive ($< 20$\%) and disruptive ($\leq 40$\%) illustrated in figure~\ref{fig:Wrad_N_D2}(b) is explained. 
For 100\%~$\textrm{D}_2$ SPI induced disruptions, the radiation at fragment arrival and the disruption processes (TQ, VDE, etc.) are of comparable levels.
The stochastisation of the field lines during the TQ might lead to the additional radiation as more heat from the core region is transported into the scrape-off layer~(SOL) plasma, where a higher (background) impurity concentration is present, consequently increasing the radiation. Additionally, higher heat loads during the pre-TQ, TQ and CQ/VDE phase could lead to increased impurity release from the wall, increasing the radiated power.
As a consequence, it is important to keep in mind, that while the radiated energy (fraction) is often used as a first indicator of the mitigation efficiency, it does not fully reflect the strain on the components induced by thermal loads, currents or forces. Despite 40\% of the energy being radiated for 100\%~$\textrm{D}_2$ SPI causing a disruption, thermal loads might still be high.

\subsection{Impact of the shatter head geometry on \frad \label{sec:Effect_geometry}}

The aim of this section is to compare the effect of the different shatter head geometries and fragment size distributions on the radiated energy fraction.
The fragment size is heavily influenced by the normal velocity in the shatter head of angle~$\alpha$: $v_\perp = v\cdot \sin (\alpha)$, where higher normal velocities typically lead to smaller mean fragment size~\cite{Parks_2016_model, Gebhart_2020_IEEE, Peherstorfer_2022_fragmentation}.
To illustrate the effects, we start by deriving a heuristic fit function for \frad in the following. We use the simple definition

\begin{equation}\label{eq:frad_thFW}
        \frad = \frac{\text{P}_\text{rad}}{\text{P}_\text{rad} + \text{P}_\text{con}} = \frac{1}{1+x},
\end{equation}

with the thermal energy flux to the wall~$\text{P}_\text{con}$ and consequently $x = \text{P}_\text{con}/\text{P}_\text{rad}$. Hereby, $\text{P}_\text{con}$ refers to the total power going to the wall from ions and electrons, accounting for their kinetic
and thermal energy (convected \& conducted; compare $\textrm{W}_\textrm{con}$ in figure~\ref{fig:energy_balance}). Physics of the plasma sheath at the plasma-wall interface reflect the following scaling for the heat fluxes onto the plasma facing components~\cite{Stangeby_2000_plasma, Artola_2021}:

\begin{equation}\label{eq:P_thFW}
        \text{P}_\text{con} \propto \text{n}_\text{e} \cdot \text{T}_\text{e}^{3/2},
\end{equation}

and from atomic physics~\cite{Hu_2021}

\begin{equation}\label{eq:P_rad}
        \text{P}_\text{rad} \propto \text{n}_\text{e} \cdot \text{n}_\text{imp} \cdot \text{L}_\text{rad}(\text{T}_\text{e}),
\end{equation}

with the electron and impurity densities ($\text{n}_\text{e}$ and $\text{n}_\text{imp}$), electron temperature~$\text{T}_\text{e}$ and the electron temperature dependent radiation factor~$\text{L}_\text{rad}(\text{T}_\text{e})$, assuming coronal equilibrium for the charge states for simplicity.

Therefore, the fraction~$x$ from equation~\eqref{eq:frad_thFW} is proportional to the inverse of the impurity density

\begin{equation}\label{eq:x_thFW}
        x \propto \frac{\text{T}_\text{e}^{3/2}}{\text{L}_\text{rad}(\text{T}_\text{e})} \cdot \frac{1}{\text{n}_\text{imp}} = \text{G}_\text{rad}(\text{T}_\text{e}) \cdot \frac{1}{\text{n}_\text{imp}}
\end{equation}

with the temperature dependent factor $\text{G}_\text{rad}$ which contains the radiation factor $\text{L}_\text{rad}$.
The impurity density itself is proportional to 

\begin{equation}\label{eq:n_imp}
        \text{n}_\text{imp} \propto \text{N}_\text{assimilated neon} = \texttt{b} / (1 + \texttt{b}/\text{N}_\text{injected neon}),
\end{equation}

with the neon assimilation parameter \texttt{b} and the total number of assimilated/injected neon atoms $\text{N}_\text{assimilated neon}$ and $\text{N}_\text{injected neon}$. Inserting this into the equation~\eqref{eq:frad_thFW}, we arrive at

\begin{equation}\label{eq:frad_fit}
        \frad = \frac{1}{1 + \frac{\text{B} \cdot \text{G}_\text{rad}(\text{T}_\text{e})}{\text{N}_\text{assimilated neon}}} = \frac{1}{1 + \frac{\texttt{a} (1 + (\texttt{b} / \text{N}_\text{injected neon}))}{\texttt{b}}}
\end{equation}

with a constant $B$ (containing all proportionality factors) or the radiation fit parameter \mbox{$\texttt{a} = \text{B} \cdot \text{G}_\text{rad}(\text{T}_\text{e})$} as well as the assimilation fit parameter \texttt{b}. While the fit parameter~\texttt{a} includes the temperature dependent radiation properties, parameter~\texttt{b} expresses the theoretical maximum number of impurity atoms a given target plasma could assimilate in the case of optimal delivery. As the plasma has a finite energy, it cannot ablate an arbitrary amount of material from the fragments. Thus there must be a limit on the assimilated atoms within the plasma, which is resembled by \texttt{b}. Also there is a trade-off between TQ duration and the time it takes for the shards to go through the plasma, the shards can only ablate during this time with a given ablation rate. Note, the parameters~\texttt{a} and~\texttt{b} do not span an orthogonal basis, hence by keeping e.g.~\texttt{b} as a fixed parameter, the open fit parameter~\texttt{a} can be used to compare the fit more quantitatively. Therefore, the fit parameter~\texttt{b} is \textbf{not} necessarily equal to the number of assimilated atoms directly.

\begin{figure}
	\centering
	\includegraphics[width=\linewidth]{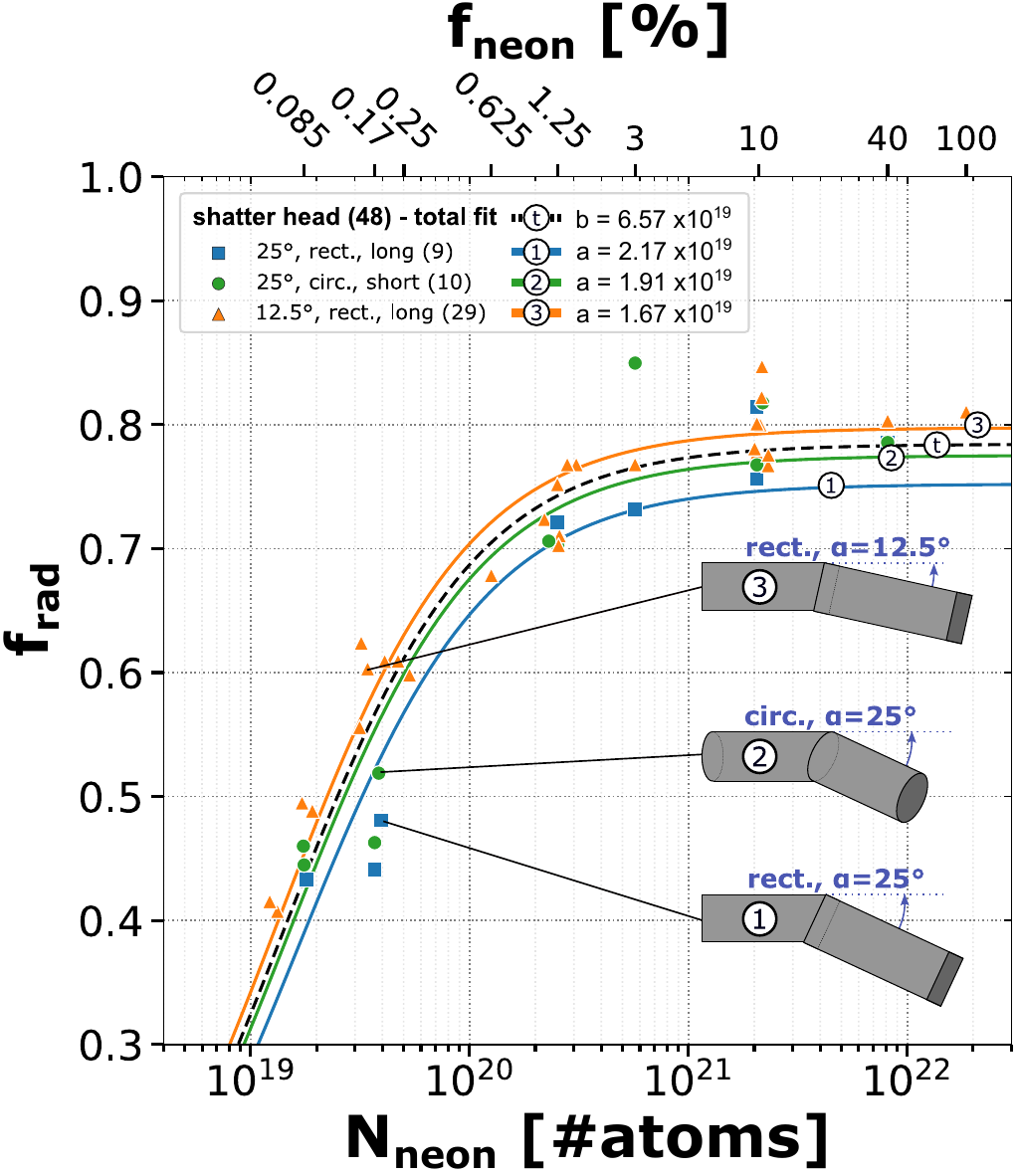}
	\caption{Radiated energy fraction as a function of the number of injected neon atoms for 8~mm diameter pellets injected into SPI~\mbox{H-modes}. The different shatter head geometries are fitted with the equation~\eqref{eq:frad_fit}. From the fit over all shatter head geometries (black dashed line) the assimilation fit \mbox{parameter~$\texttt{b} = 6.57\times 10^{19}$} was used for the other fits as a fixed parameter for better comparability. The fit parameters are given in table~\ref{tab:fit_parameters}. The three highlighted scatter points connected to the respective heads shared similar pre-shattering parameters of the pellets ($v_\textrm{pre-shattering} \sim 430$--$460$~m/s, 0.17\% neon). The shallow angle $12.5^{\circ}$, rectangle shatter shows the overall highest \frad-values. The largest impact of the shatter geometry/normal velocity is observed in the $10^{19}$--$10^{20}$ injected neon atom range~(0.085\% and 0.17\% for 8~mm -- neon doping to suppress plasmoid drift).}
	\label{fig:shatter_head_fits}
\end{figure}

For 8~mm diameter pellets injected into the SPI \mbox{H-mode} plasmas, figure~\ref{fig:shatter_head_fits} shows the effect of the shatter head geometry with the fit function from equation~\eqref{eq:frad_fit}. The fits were created with the \texttt{scipy.optimize.curve\_fit} function and the error estimate via \texttt{np.sqrt(np.diag(pcov))} as suggested in the \texttt{scipy} documentation~\cite{scipy_curve_fit}.
In table~\ref{tab:fit_parameters} the fitting parameters from equation~\eqref{eq:frad_fit} (including errors) are given.
Hereby, the assimilation parameter~\texttt{b} was fixed to allow a better comparison between the shatter heads as the fit parameters are not orthogonal.
The fixed factor \mbox{$\texttt{b} = 6.57 \times 10^{19}$} was determined by the total fit~\circled{t} for all three shatter head geometries (dashed black line in figure~\ref{fig:shatter_head_fits}).
Hereby, lower values of~\texttt{a} will result in higher values of \frad, hence are considered optimal.

\begin{table*}[htb]
        \caption{Fit parameters and least squares residual for each fit from figure~\ref{fig:shatter_head_fits}. Hereby, smaller values for \texttt{a} (for fixed \texttt{b}) lead to higher \frad-values. The fitting and error estimate was done via the \texttt{scipy.optimize.curve\_fit} with the error estimate as \texttt{np.sqrt(np.diag(pcov))}~\cite{scipy_curve_fit}.}
        \label{tab:fit_parameters}
        \centering
        \def\arraystretch{1.5}
        \begin{tikzpicture}
            \node (table) [inner sep=0pt] {
            \begin{tabular}{l|l|l|l|l}
                fit number & \texttt{a} param. & \texttt{a} error (in \%)& \texttt{b} param. & \texttt{b} error (in \%)\\
                \hline
                \hline
                \multicolumn{5}{l}{figure~\ref{fig:shatter_head_fits}(a) -- shatter geometry effect -- $^{\ddagger}$ fixed parameter}\\
                \hline
                \circled{t} (total; black) & $1.81 \times 10^{19}$ & $1.23 \times 10^{18}$ (6.8\%) & $6.57 \times 10^{19}$ & $6.70 \times 10^{18}$ (10.2\%)\\
                \hline
                \textcolor{pythonblue}{\circled{\textcolor{black}{1}}} ($25^{\circ}$ rect.) & $2.17 \times 10^{19}$ & $1.54 \times 10^{18}$ (7.1\%) & $6.57 \times 10^{19 \ddagger}$ & \multicolumn{1}{c}{--} \\
                \hline
                \textcolor{pythongreen}{\circled{\textcolor{black}{2}}} ($25^{\circ}$ circ.) & $1.91 \times 10^{19}$ & $1.50 \times 10^{18}$ (7.9\%) & $6.57 \times 10^{19 \ddagger}$ & \multicolumn{1}{c}{--} \\
                \hline
                \textcolor{pythonorange}{\circled{\textcolor{black}{3}}} ($12.5^{\circ}$ rect.) & $1.67 \times 10^{19}$ & $4.79 \times 10^{17}$ (2.9\%) & $6.57 \times 10^{19 \ddagger}$ & \multicolumn{1}{c}{--}
            \end{tabular}
            };
            \draw [rounded corners=.5em] (table.north west) rectangle (table.south east);
	\end{tikzpicture}
\end{table*}

\begin{figure}
	\centering
	\includegraphics[width=\linewidth]{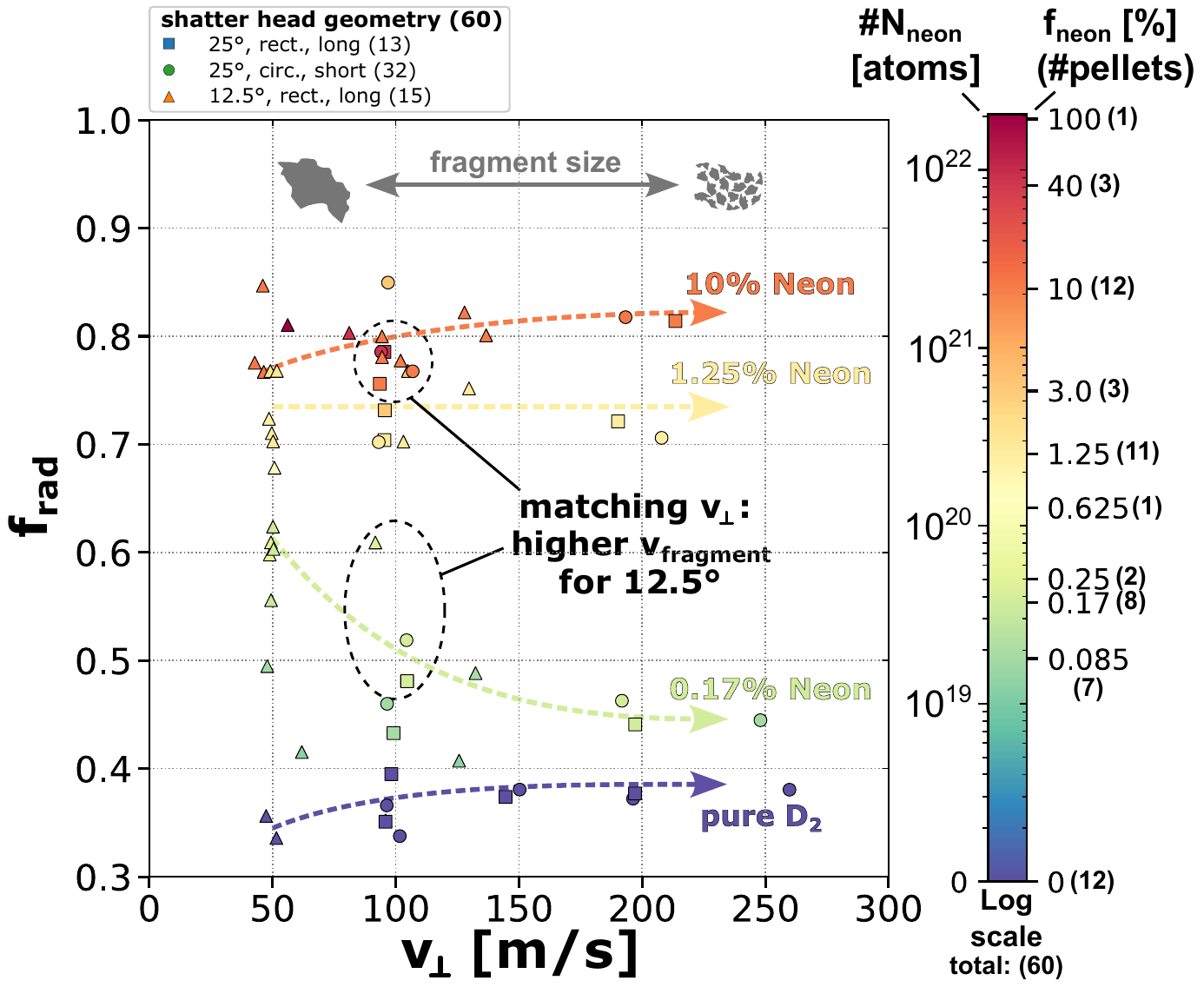}
	\caption{Radiated energy fraction as a function of the perpendicular velocity component (proxy for fragment size). Only 8~mm diameter, single, disruptive injections into the SPI H-mode are shown which did not trigger the \VPEState-change of the DCS before the IP-spike. The colour-coding indicates the estimated number of neon atoms. For the 0.17\% neon pellet, the arrow indicates decreasing values of frad for smaller fragments, which is in line with figure~\ref{fig:shatter_head_fits}. For higher neon fractions, a potential impact of the fragment size on the radiated energy fraction is within the measurement uncertainties of 10\%.}
	\label{fig:frad_trends}
\end{figure}

For high neon content (above $10^{21}$ neon atoms; $\textrm{f}_\textrm{neon} \geq 10\%$ neon), the shatter head geometry -- and therefore the fragment size and velocity distribution -- seems to only play a minor role (2--5\% difference; within the uncertainties estimated in section~\ref{sec:Wrad}) as shown in figure~\ref{fig:shatter_head_fits}. It shows a larger impact (10--20\% difference) on injections with small neon doping (around 0.085--1\% neon for the 8~mm pellets i.e. $10^{19}$--$10^{20}$ neon atoms) which were performed to investigate the effects of low neon doping on the plasmoid drift suppression. Hereby, the $12.5^{\circ}$ rectangular shatter head -- producing the larger and faster fragments compared to its $25^{\circ}$ counter part -- is at the top of the~\frad distribution. The lowest radiated energy values are observed for the $25^{\circ}$ rectangular shatter head. The short, $25^{\circ}$ circular shatter head is mostly in between these two and the \frad-value varies strongly, as its effective shatter angle is a function of pellet impact position inside the shatter head~\cite{Peherstorfer_2022_fragmentation} shown in figure~\ref{fig:head_comparison}(d). This is also in line with the results for 100\%~$\textrm{D}_2$ SPI presented by \mbox{Jachmich~\etal~\cite{Jachmich_2023_EPS, Jachmich_2024_EPS}} for the material assimilation -- which is crucial to increase the free electron density in the core to suppress runaway electron generation for future machines. Hereby, shallower shatter geometries seem beneficial for material assimilation. We observe the strongest impact of the SPI parameters (head geometry, pre-shattering pellet velocity, neon content, ...) on the disruption behaviour (i.e. frad, CQ dynamics, ...) for the low neon doping not only in the experiments~\cite{Heinrich_2023_EPS, Heinrich_2025_PhD} but also DREAM simulations~\cite{Halldestam_2024_REM, Halldestam_2024}. The cases with \mbox{$\leq 10^{20}$} neon atoms in the pellet seem the most sensible to injection parameter changes. Probably, the reason for this is that for 100\%~$\textrm{D}_2$ injection the radiation from the pellet materials is complemented by other, ill-controlled factors (such as wall conditions or impurities from the vessel), whereas above a critical number of neon atoms the neon line radiation dominates over the impact of different fragment delivery parameters.
Another potential explanation is the observed change of inside-out to outside-in cooling, which was observed for this parameter range in INDEX simulations by A.~Patel~\cite{Patel_2024}.
For higher neon content, a cold front develops that lags slightly behind the pellets, causing an outside-in cooling of the plasma. Here, a variation in fragment size and velocity may cause different cooling/radiation properties.

Figure~\ref{fig:frad_trends} shows the radiated energy fraction as a function of the normal velocity component~\vperp for the different neon concentrations used as colour-coding. 
Similar to the observed trend in figure~\ref{fig:shatter_head_fits}, we can observe at low neon content -- indicated as 0.17\% in the figure -- that lower values of \vperp (i.e. larger fragments) lead to higher \frad-values. This is in line with the findings for material assimilation~\cite{Jachmich_2023_EPS, Jachmich_2024_EPS}, hence to maximize the assimilation of deuterium for ITER, the present SPI design choice foresees a $15.5^{\circ}$ angle for the shattering unit~\cite{Lehnen_2023_FEC, Kruezi_2024_3rdTM}. 

Around 1.25\% neon, frad seems to be independent of \vperp, illustrated by the horizontal, yellow line in figure~\ref{fig:shatter_head_fits}.

For high neon content (10\% neon, \mbox{$2 \times 10^{21}$} neon atoms) -- where \frad saturates -- the difference in \frad is found within the uncertainty estimate, and thus a significant trend has not been observed. A minor benefit of smaller fragments may be visible in the experiments~\cite{Heinrich_2024_3rdTM, Heinrich_2024_APS, Heinrich_2025_PhD} (see also figure~\ref{fig:frad_trends}) and AUG SPI JOREK simulations~\cite{Tang_2024, Tang_2024_AAPPS}. There are relatively few data points available for the $25^{\circ}$ shatter heads with 10\% neon pellets for a direct comparison to the $12.5^{\circ}$ head, thus further experiments will be necessary to verify a potential change in trend from low to high neon content.

At present, this apparent change in trend is not yet fully understood. However, we want to highlight two factors which may be contributing to this:
\begin{itemize}
	\item The potential change of inside-out towards outside-in cooling may take place in the low neon doping range~\cite{Patel_2024}, which could lead to different plasma radiation properties after the injections, strongly depending on the delivery of the material: large or small, slow or fast fragments and the degree of plasmoid drift suppression.
	\item The fragment size is a non-linear function of \vperp as observed experimentally for the ASDEX Upgrade SPI system~\cite{Peherstorfer_2022_fragmentation}. For high perpendicular velocities, the mean fragment size does not vary as much as the mean fragment velocity. This could cause a potential roll over effect, where the increased mean fragment velocity outweighs the reduced size.
\end{itemize}

In the case, the low neon doping would not be sufficient for a strong reduction in the plasmoid drift, the larger~\& faster fragments in the low neon range could lead to a better deposition of the material in the core region, hence increased material assimilation~\cite{Jachmich_2023_EPS, Jachmich_2024_EPS} and radiation (see figures~\ref{fig:shatter_head_fits} and~\ref{fig:frad_trends}). For high neon content, the smaller fragments may enhance the outside-in cooling, however, further studies on the radiation effectiveness and \frad are currently prepared for the ASDEX Upgrade INDEX simulations.

\section{Summary \label{sec:Summary}} 
In support for the ITER DMS, a highly flexible, triple-barrel SPI system was installed at the tokamak ASDEX Upgrade~\cite{Dibon_2023_SPI, Jachmich_2023_EPS}.

Based on experimental results at ASDEX Upgrade~\cite{Sheikh_2020_AUG} as well as JOREK simulations for AUG VDEs~\cite{Schwarz_2024_PhD}, the coupled energy fraction \fc was estimated to 50\%, necessary to calculate the radiated energy fraction~\frad.
The total radiated energy~\Wrad and radiated energy fraction~\frad are a strong function of the neon content inside the pellet.
While non-disruptive SPI shows radiated energy fractions of \mbox{$\leq 20$\%}, the fraction increases to a total of \mbox{$\leq 40$\%} after the TQ and VDE phase of disruptive injections. Already with small neon doping of the pellets ($10^{19}$--$10^{20}$ neon atoms, equivalent to 0.085\%--1\% neon inside 8~mm diameter pellets) -- to study the ablation physics and plasmoid drift suppression -- \frad-values between 40--70\% are observed. Increasing the neon content even further, the \frad-curve saturates around 80\%.
Overall, SPI with the $12.5^{\circ}$, rectangular shatter head -- producing large and fast fragments -- caused the highest radiated energy fractions.
The largest effect of the shatter head geometry was observed for low neon doping of \mbox{$\approx$ 0.085--0.17\%} in the 8~mm diameter pellets (\mbox{$\lesssim 2 \times 10^{19}$} neon atoms), with an increase in radiated energy in the 10--20\% range. Largest sensitivity to injection parameters was also found for this amount of neon doping in recent DREAM simulations~\cite{Halldestam_2024_REM, Halldestam_2024}.
The effect of the shatter head geometry on the radiated energy fraction for high neon concentrations (above \mbox{$2 \times 10^{21}$} neon atoms or 10\% neon 8~mm pellets) was observed to be in the 2--5\% range, which is within the present uncertainty estimate of inter-shot comparisons.
Comparing injections of different normal velocity components (proxy for fragment size), large fragments seem beneficial for \frad in the case of neon doped pellets (below approximately \mbox{$2\times 10^{20}$} neon atoms), while a clear trend can not be proven with the current uncertainties for higher neon concentrations.
We observed that full-size, 8~mm diameter, 100\%~$\textrm{D}_2$ pellets are sufficient to cause the H-mode plasmas to disrupt. In contrast, plasmas with injections of 4~mm pellets or shortened 8~mm pellets (with about half the nominal length) can typically recover after the injection. No dedicated, fine scan of the amount of injected deuterium to trigger the disruption had been performed, however, the threshold is roughly located around \mbox{$1\times 10^{22}$} deuterium molecules in line with previous simulations results~\cite{Hoelzl_2020_D2-SPI}. 

\paragraph{Acknowledgements}
\footnotesize

\noindent The authors are grateful to P.~David for his work on the bolometry system upgrade and the measurements; M.~Maraschek, M.~H\"olzl, P.~Halldestam, A.~Patel and W.~Tang for the fruitful discussions; and to A.~Bock, R.~Schramm, R.~Fischer, and B.~Kurzan for their help during the SPI experiments.
This work has been carried out within the framework of the EUROfusion Consortium, funded by the European Union via the Euratom Research and Training Programme (Grant Agreement No 101052200 — EUROfusion). Views and opinions expressed are however those of the author(s) only and do not necessarily reflect those of the European Union, the European Commission, or the ITER Organization. Neither the European Union nor the European Commission can be held responsible for them.
This work receives funding from the ITER Organization under contract IO/20/IA/43-2200. The ASDEX-Upgrade SPI project has been implemented as part of the ITER DMS Task Force programme. The SPI system and related diagnostics have received funding from the ITER Organization under contracts IO/20/CT/43-2084, IO/20/CT/43-2115, IO/20/CT/43-2116.

\end{document}